\documentclass{aastex63}

\usepackage{amssymb}
\usepackage[T1]{fontenc}

\def\eg{{\it e.g.,} }
\def\etal{{\it et al. }}
\def\ie{{\it i.e.,} }
\def\rc{r_{\rm c}}
\def\rg{r_{\rm g}}
\def\rs{r_{\rm s}}
\def\mbh{M_{\rm {bh}}}
\def\vr{v}

\def\vp{v_{\rm \phi}}

\def\thvir{\Theta_{\rm vir}}
\def\Thmx{\Theta_{\rm max}}
\def\Thob{\Theta_{\rm ob}}
\def\rh{r_{\rm h}}

\def\vk{v_{\rm K}}
\def\lmk{\lambda_{\rm K}}
\def\lmdob{\lambda_{\rm ob}}

\def\lsim{\lower.5ex\hbox{$\; \buildrel < \over \sim \;$}}
\def\gsim{\lower.5ex\hbox{$\; \buildrel > \over \sim \;$}}
\def\simeq{\lower.3ex\hbox{$\; \buildrel \sim \over - \;$}}
\def\rin{r_{\rm in}}
\def\rout{r_{\rm ob}}
\def\as{a_{\rm s}}

\def\sun{M_\odot}

\def\rin{r_{\rm in}}
\def\tt{\tilde{t}}

\def\be{B_{\rm ob}}

\def\hm{{\it hot-mode} }
\def\cm{{\it cold-mode} }
\submitjournal{ApJ}

\shorttitle{General advective accretion disks with the two-mode gas inflows}
\shortauthors{Kumar \& Yuan}

\begin{document}
\title{Review the possible advective disk structures around a black hole with two-type gas inflows}
\email{rkumar@ustc.edu.cn; yfyuan@ustc.edu.cn}
\author{Rajiv Kumar}
\affil{CAS Key Laboratory for Research in Galaxies and Cosmology, Department of Astronomy, University of Science \& Technology of China, Hefei,
Anhui 230026, China}

\author{Ye-Fei Yuan}
\affiliation{CAS Key Laboratory for Research in Galaxies and Cosmology, Department of Astronomy, University of Science \& Technology of China, Hefei,
Anhui 230026, China}
\affil{School of Astronomy and Space Sciences, University of Science \& Technology of China, Hefei 230026, China}



\begin{abstract}
We studied the general advective accretion solutions around the Kerr black hole (BH) with investigating two types of inflow gases at the outer accretion boundary (AB). We classified these two types of gases as a \cm and a \hm inflow gas at the outer AB on the basis of their temperatures and solutions. We found that the \hm gas is more efficient for angular momentum transportation around the outer AB than the \cm gas.
The \hm gas can give global multiple \cite[popular as shock solution][]{c89} or single sonic point solutions and the \cm can give smooth global solution \cite[popular as ADAF][]{ny94} or two sonic point solutions.     
These solutions also represented on a plane of energy and angular momentum ($\be-L_0$) parameter space. Theoretically for the first time, we explored the relation between the nature of accretion solutions with the nature of initial accreting gas at the AB with detail computational and possible physical analysis.  
 We also found that the surface density of the flow is highly affected with changing of the temperature at the AB, which can alter the radiative emissivities of the flow.  
 The flow variables of various advective solutions also compared.
On the basis of those results, we plotted some inner disk structures around the BHs. Doing so, we conjectured about the persistent/transient nature of spectral states, soft-excess and time scales of variabilities around the black hole $X-$ ray binaries (BXBs) and active galactic nuclei (AGNs).    
\end{abstract}
\section{Introduction} \label{sec:intro}
The black holes are gravitationally most powerful objects in the Universal but their presence is only depicted by the external environments, in other words, 
their strong power is nothing without the presence of others, for instance, external light, other celestial objects, external gases (from companion stars, ISM, 
and torus gas). The accreting BHs are observed in all band of the electromagnetic spectrum and exhibited many types of phenomena, \eg various spectral states 
and their transitions, bipolar outflows/jets, quasi-periodic oscillations (QPOs) and so on \citep{gfp03,fbg04,mkkf06,detal12,m12}. The active BHs are mostly found in the center of galaxies, known as  the
AGNs and as the primary object of the many binary systems, known as the BXBs. 
Since the BHs are powered by the accretion disk and their accretion processes, therefore many studies 
have done on the accretion flows as well as structures of the disk. The disk formation needs the rotation, so the first theoretical model with rotation and viscosity in the flow 
was given by \cite{ss73} \cite[a relativistic version by][]{nt73} and known as the Keplerian disk (KD) model. This model is basically non-advective and ad-hoc in nature, which explains the thermal components of the BXBs and AGNs.         

The basic tendancy of an angular momentum (AM) is that to make an object or gas in an equilibrium with the gravity along/around an equatorial plane of the central object. In the accretion physics, the AM distribution of the gas is termed as the Keplerian, super-Keplerian and sub-Keplerian flows, when it compared with a centrifugal force and gravity force. If the viscous accretion flow is having the Keplerian AM distribution then the both forces are in the equilibrium therefore the disk flow is almost non-advective and thermalized, \eg  KD model. 
The accretion flow with the sub-Keplerian AM distribution can have more exotic nature because of mismatch of the  both forces and this situation can lead to introduce more kind of effective local forces in the fluid flow. So this situation can create strong bulk velocity (bulk momentum gradient force) 
and thermal pressure (pressure gradient force) in the disk, which can alter fastly/slowly dynamical and physical processes, like, viscous, advection and cooling processes, and that can produce variabilities, non-thermal radiations, ejection of energetic gas particles and so on. 
Moreover, the sub-Keplerian accretion flow is also more relevant for the understanding of the global transonic accretion flows, so there are many studies have done with it in more than last four decades \citep{i77,f87,c89,ny94}. 
The sub-Keplerian flow can give many types of the advective solutions depend on the variation of the AM distributions in the flow. Two types of the advective solutions for the hot accretion flow are very popular in the literature, one, the shocked accretion flow \citep{f87,c89,msc96b,lmc98,m03,gl04,bdl08,dc12,gc13,dcnm14,lckhr16,kc17,ddmn19,lb20,gbc20} and the other, a special kind of the smooth solution known as the Advection-Dominated Accretion Flow (ADAF) solution \citep{i77,ny94,h96,nkh97,lgy99,kfm08,nsp12,yn14,kg18,kg19}.  
Interestingly, both the advective disk solutions/models can be generated from the same set of the fluid differential equations and assumptions, but 
both (shock and ADAF) solutions have basic difference of AM distribution. So here we will explore one of possible reason of the different AM distributions with distinguishing properties of the accreting gas at the AB.

As we know that the BXBs are typically divided into two types on the basis of their companion star \cite[][and references therein]{nb18}. So, if the companion star of the BXB is in high mass category ($>10\sun$) then known as high-mass BH $X-$ray binary (HMBXB). If it is in low mass category ($<3\sun$) then known as low-mass BH $X-$ray binary (LMBXB). In between these two mass gap of the companion stars in the BXBs can also be termed as an intermediate-mass BH $X-$ray binary (IMBXB). Interestingly, LMBXBs are usually known as the transient $X-$ray sources and HMBXBs are known as the persistent $X-$ray sources \citep{nb18}. The possible reasons of this difference in both may be come due to their different feeding mechanism, geometry of the system, nature and abundance of the accreting gas, and so on. For instance, the LMBXBs are undergo Roche lobe overflow and the BH is mainly fed by the over flow of the companion star,  
and BH of the HMBXBs are mainly fed by the winds of the companion. 
So, we believed that the temperature of the accreting gas can be different at the AB, like, a neutral gas or partially/fully ionized gas \citep{s73a}, and it may also be dependent on the feeding mechanisms, like, the Roche lobe flow, stellar winds or failed winds from the disk. Similarly, the AGNs are also having different nature of the accreting gas sources and feeding mechanisms, like, the torus gas (mostly neutral), ISM, winds and cold clumpy clouds \cite[it maybe mixture of the cold gas surrounded by hot dilute gas][]{byz20} and so on. 
So, we are interested to investigate the advective accretion solutions with the different temperatures of the accreting gas at the AB, and expected to focus the some light on the different behaviours of the accreting BHs.

The BXBs are showing the many types of the spectral states and their state transitions with exhibiting many phenomena, like, on/off of the bipolar jets, evolutions of the QPOs and so on \citep{gfp03,m12,nb18}. So,  
there are two most prevailing hybrid disk models to understand the spectral state variations in the accreting BHs, one, the two-zone radial accretion theory \citep{nym95,h96,emn97,mfp18}, and the other, the sandwiched geometry disk known as the two component accretion flow \citep{ct95,mc10,gc13}. Both models have the KD common and sub-Keplerian flow with smooth or shocked flow but their geometrical orientation is different.  
In the present article, we have predicted possible combinations of the hybrid disk structures with combining both the models on the basis of the obtained theoretical results with two types of accreting gas at the AB. Previously, there are many configurations of the hybrid disk structures have also been predicted by \cite{wl91}. Here, some of them we will discuss on the basis of the advective solutions. The present work is inspired from the recent 2D study \citep{kg18}, which has shown the shock like behaviors above the equatorial plane with the ADAF solutions on the equator.

Moreover, there are some parameters and properties of the accreting gas, which can help in the understanding of the observed properties of the accreting BHs:
A) {\it BH characteristics}- 1. Mass, 2. Spin, 
B) {\it External environment}- 
3. System geometry and mode of gas feeding sources, e.g., binary separation, binary period, eccentricity, Roche-lobe over flow, stellar wind, ISM gas, and so on, 
4. Abundance and nature of accreting gas, e.g., temperature of the gas, non-magnetised or magnetised gas \citep{s73b} (it may depend on the low/high mass of the donor star in case of the BXBs!), gas composition/metallicity  and so on, 
C) {\it Inner disk environment}- 
5. Inner flow geometry with/without shock, and 
6. Magnetic field structure and strength.
Here, the A option is inherent since invariant for very long time scale, but the C option can mostly depend on the option B. 
So we need to more explore the B option via the observations and theories. 
The possible importance of these properties of the accreting BHs are, 
1. Mass- can help to understand the size and time scales of evolutions,
2. Spin- can help to understand the strength of relativistic effects,
3. 
System geometry and mode of gas feeding sources- may help to understand the disk size, 
timing of the gas availability, abundance  of the accreting gas, initial dynamics of the gas and so on,
4. Abundance and nature of accreting gas- may help to determine the inner/outer disk structures with nature of the advective/non-advective flows, spectral state dominance and their transitions, generation of the bipolar outflows, and
accumulation of magnetic field and so on, which  can help in the jet generation, acceleration and collimation.
Peoples are investigating of those parameters and the properties of the gases with the help of available observations and theories, e.g.,  the multi-wavelength EM radiation studies (spectral radiation with emission mechanisms, reflection/polarization studies and BH imaging) and GW observations (collision of compact objects). 
Apart from many studies to understand the accreting BHs, there are last two options (B and C) are need to explore more with the theories and observations. 
In the present theoretical study, we have explored the inner structure of the disk with nature of the accreting gas at the outer AB with some assumptions (Both options B and C). In the present model equations, we have used two things different from the other previous many studies of the accretion flows: One, we have used variable $\Gamma$ EoS \citep{cr09} for viscous flow around the Kerr BH, 
and second, we have used full relativistic $r-\phi$ component of the viscous stress tensor, unlike to the other simplified forms have been used by many authors \citep{nt73,acg96,pa97,s09,ck16}.   

The goal of the present study is to investigate the general advective solutions with two mode of the accreting gas, like, the ADAFs, smooth solutions, shock solutions, etc 
and represent them on a single plane of parameter space and relate them with possible physical situations or events in the astronomy. 
The structure of the paper is in next section \ref{sec:eqns}, the relativistic fluid equations and assumptions; section \ref{sec:nmethod}, numerical method;
section \ref{sec:result}, results and in the final section, summary and discussions.

\section{Relativistic fluid Equations and Assumptions} \label{sec:eqns}
We considered steady-state advective viscous general relativistic hydrodynamic (GRHD) fluid equations with axisymmetric in Boyer-Lindquist
coordinates \citep{bl67} around the Kerr BH. We also assumed that the accreting matter has the AM and following vertical equilibrium along the $z$- direction. 
Thus the accreting gas formed rotating disk around the equatorial plane. Although, the vertical equilibrium condition can not be correct very close to the BH, when the outflows occurred otherwise the disk has equilibrium. 
For time being, we have ignored outflows as well as magnetic field in this study.
The relativistic equations and flow variables are represented in the geometrical unit
and chosen $G=c=\mbh=1$ otherwise stated, where, $G$, $\mbh$ and $c$ are the universal
gravitational constant, mass of the BH, and the speed of light, respectively.
We used the Kerr background geometry in cylindrical coordinates \citep{nt73} and can be written on the equatorial plane as,
\begin{equation}
ds^2=g_{tt}dt^2+2g_{t\phi}dtd\phi+g_{rr}dr^2+g_{\phi\phi}d\phi^2+dz^2,
\end{equation} 
where $g_{tt}={\cal A}w^2/r^2-r^2\Delta/{\cal A}$, $g_{t\phi}=-w{\cal A}/r^2$, $g_{rr}=r^2/\Delta$ and $g_{\phi\phi}={\cal A}/r^2$. 
Here $\Delta=r^2-2r+a^2$, $w=2ar/{\cal A}$, ${\cal A}=r^4+r^2a^2+2ra^2$ and $a$ is the kerr spin parameter. 
Following \cite{kc17,ck16}, the conservative form of GRHD fluid equations can be written as
\begin{equation}
T^{\mu\nu}_{;\nu}=0, ~~~\mbox{and}~~~  (\rho u^\nu)_{;\nu}=0,
\label{cgr.eq}
\end{equation}
where $T^{\mu\nu}=(e+p)u^\mu u^\nu+pg^{\mu\nu}+\tau^{\mu\nu}$ is the energy-momentum tensor of the RHD fluid, $\tau^{\mu\nu}$ is viscous stress tensor, and $g^{\mu\nu}$ is Kerr-metric tensor and indices $\mu$, $\nu$ represent the space-time coordinates ($t,r,\phi,z$). The
$\tau^{\mu\nu}=-2\eta\sigma^{\mu\nu}$, where $\sigma^{\mu\nu}$ is shear tensor and $\eta=\rho\nu$ is a viscosity coefficient. Here, $\nu=\alpha \as r f_c$ and $f_c=1/\gamma_v^2$ with following \cite{pa97} and $\alpha, \as$ and $\gamma_v$ are a viscosity parameter, sound speed and bulk Lorentz factor of the flow, respectively.
Here, $\rho$, $p$, $e$, and $u$'s are the local gas density, local gas pressure, local energy density and four velocities of the flow, respectively. 
Projecting the conserve form of the energy-momentum equation of (\ref{cgr.eq}) along the spatial $i$th direction \ie $h_{\mu}^{i}T^{\mu\nu}_{;\nu}=0$ and can be re-written as
\begin{equation}
(e+p)u^{\nu}u_{;\nu}^i+(u^{i}u^{\nu}+g^{i\nu})p_{,\nu}+h^{i}_{\nu}\tau^{\mu\nu}_{;\nu}=0.
\label{NS.eq}
\end{equation}
This equation is known as the relativistic Navier-Stokes equation for viscous flow and $h^{i}_{\nu}=u^{i}u_{\nu}+g^{i}_\nu$ is the projection tensor ($i=r, \phi, z$).
Thus this equation has three components in three spatial directions, which are represented below with following some disk assumption, \eg axisymmetric, vertical equilibrium along $z$- direction and 
we also assumed only $r-\phi$ component of the viscous stress tensor is non-zero.  Now the simplified components of the equation (\ref{NS.eq}) are $r-$ component momentum balance equation, 
\begin{equation}
u_r u^r_{,r}+\frac{1}{2}[g_{rr,r}(u^r)^2-g_{tt,r}(u^t)^2-2g_{t\phi,r}u^tu^\phi-g_{\phi\phi,r}(u^\phi)^2]+(1+u_ru^r)\frac{p_{,r}}{e+p}=0,
\label{rNS.eq}
\end{equation} 
$\phi-$component equation, 
\begin{equation}
\tau^r_\phi=\rho u^r(L-L_0)=-2\eta\sigma^r_\phi,
\label{pNS.eq}
\end{equation}
and $z-$component with vertical equilibrium condition gives disk half-height prescription with following \cite{rh95,pa97}, 
\begin{equation}
H=rH_\theta=r\sqrt{\frac{p}{\rho}\frac{r}{F_e}} ,
\label{hh.eq}
\end{equation}
where, $F_e=\gamma_\phi^2[(r^2+a^2)^2+2\Delta a^2]/[(r^2+a^2)^2-2\Delta a^2]$, $\gamma_\phi=1/\sqrt{1-\Omega\lambda}$ is bulk azimuthal Lorentz factor, $\lambda=-u_\phi/u_t$ is the specific AM of the flow, $\Omega=u^\phi/u^t$ is the angular frequency of the flow, $L=hu_\phi$ is bulk AM of the fluid flow, and $L_0$ is bulk AM at the horizon. 
The mass conservation form of equation (\ref{cgr.eq}) can be written as,
\begin{equation}
-\dot{M}=4\pi r^2H_\theta\rho u^r,
\label{mdot.eq}
\end{equation}
where $H_\theta$ is local angular scale of the disk flow and arises due to vertical averaging of $z$-component of the four-velocity $u^z \sim 0$. 

The energy generation equation or 1st law of thermodynamic is $u_\mu T^{\mu\nu}_{;\nu}=0$, and after simplification with model assumptions it can be written as,
\begin{equation}
u^r[h\rho_{,r}-e_{,r}]=q^+,
\label{reg.eq}
\end{equation}
 where, $q^+=\tau^{r\phi}\sigma_{r\phi}$ is viscous heating rate and $h=(e+p)/\rho$ is specific enthalpy of the flow. 
All the flow variables of this model equations are defined on the equatorial plane and dependent with radial distance, $r$.
Following \cite{kc17}, we used variable $\Gamma$ equation of state (EoS) of multi-species flow, which is given by \cite{cr09}, so the fluid can contained electrons, positrons and protons of different proportion with maintaining charge neutrality. Thus the internal energy density has obtained as
\begin{equation}
e=\Sigma_je_j=\frac{\rho}{\tilde{\tau}}f,
\label{eos.eq}
\end{equation}
where, $e_j$ is energy density for the particular species \cite[See details in][]{cr09}, $\rho=\Sigma_jn_jm_j=\rho_e\tilde{\tau}$ is total mass density, $\tilde{\tau}=[2-\xi(1-1/\chi)]$ and $f=(2-\xi)\left[1+\Theta(9\Theta+3)/(3\Theta+2)\right]+\xi\left[{1}/{\chi}+
\Theta(9\Theta+3/\chi)/(3\Theta+2/\chi)\right]$. Here, $\Theta=kT/(m_ec^2)$ is dimensionless temperature, $\xi=n_p/n_e$ is composition parameter of the flow and $\chi=m_e/m_p$ is mass ratio of the electron and proton. Here $n_e$ and $n_p$ are the electron and proton number densities, respectively.  $\xi\in[0,1]$, and the extreme low and high values are representing the pure electron-positron and electron-proton flow, respectively. 
In the relativity, the adiabatic index ($\Gamma$) and sound speed ($\as$) are defined as 
\begin{equation}
\Gamma=1+\frac{1}{N}; ~~N=\frac{1}{2}\frac{df}{d\Theta} ~~~\mbox{and} ~~~\as^2=\frac{\Gamma p}{e+p},
\end{equation}
where, $p=\Sigma_jp_j=2\rho\Theta/\tilde{\tau}$ is isotropic pressure for the single-temperature flow. 

Since we assumed only $r-\phi$ component of the viscous tensor is effective in $r-\phi$ plane and others are zero. Therefore, the $r-\phi$ component of the shear tensor is presented here \citep{pa97,ck16},
\begin{equation}
2\sigma^r_\phi=u^r_{;\phi}+g^{rr}u_{\phi;r}+a^ru_\phi+a_\phi u^r-\frac{2}{3}\Theta_{\rm exp}u^ru_\phi,
\label{srp.eq}
\end{equation}
where, $a$'s are the four-acceleration and $\Theta_{\rm exp}=u^\gamma_{;\gamma}$ is the expansion of fluid world line. Following the definitions of the four-acceleration and covariant derivatives of four-velocity with some suitable algebraic simplification then we can re-write the equation (\ref{srp.eq})  as	
 \begin{equation}
2\sigma^r_\phi =(g^{rr}+u^ru^r)u_{\phi,r}+\frac{1}{3}u^ru_\phi u^r_{,r}+\tilde{\sigma},
\label{srp1.eq}
\end{equation}
where $u_{\phi,r}$ and $u^r_{,r}$ are the ordinary derivatives of the four-velocities with respect to `$r$'. $\tilde{\sigma}=u^r_{;\phi}-g^{rr}\Gamma^\gamma_{\phi r}u_\gamma+u_\phi[u^r\Gamma^r_{r\gamma}u^\gamma+u^tu^r_{;t}+u^\phi u^r_{;\phi}]+u^r[u^t u_{\phi;t}-u^r\Gamma^\gamma_{\phi r}u_\gamma+u^\phi u_{\phi;\phi}]-2u^ru_\phi[u^t_{;t}+u^\phi_{;\phi}+u^z_{;z}+\Gamma^r_{r\gamma}u^\gamma]/3$, and $\Gamma^\gamma_{\mu\nu}$ is the Christoffel symbol. 

After some algebraic simplifications with using relations of the four velocity to three velocity of the flow then the equation (\ref{rNS.eq}) with integrating along the radial direction can be written as
\begin{equation}
 \int v\gamma_v^2dv+\int \frac{dP}{e+p}+\int{\cal F}dr=E^{'},
 \label{irme.eq}
\end{equation}
here ${\cal F}=[1/r+\Delta^{'}/\Delta-D^{'}/D]/2$, $\Delta^{'}=2(r-1)$, $D=r^3+(r+2)a^2-(r-2)\lambda^2-4a\lambda$, and $D^{'}=3r^2+a^2-\lambda^2$. 
By using definition of $h$ and equation (\ref{reg.eq}), the term $\int dp/(e+p)$ can be written as
\begin{equation}
\int \frac{dp}{e+p}=\int\frac{dh}{h}+\int\frac{q^+}{\rho hu^r}dr.
\label{pirme.eq}
\end{equation}
Now, replacing second term of equation (\ref{irme.eq}) by equation (\ref{pirme.eq}), and after doing some algebra, we get 
\begin{equation}	
E=h\gamma_v\sqrt{r\Delta}~{\rm exp(X_f)}
 \label{ec.eq}
\end{equation}
This is a specific energy constant of the motion in presence of the viscosity \citep{ck16}. It is a constant of motion because comes from the first principle with integration of all the equations of motion. Here, $X_f=\int[{\cal F}_p+q^+/(\rho u^rh)]dr$ and ${\cal F}_p=-D^{'}/{2D}$. By using the equation (\ref{pNS.eq}) and $\tau^r_\phi=-2\eta\sigma^r_\phi$ then the expression of $q^+$ can be written as $q^+=(g_{rr}/g_{\phi\phi})\tau^r_\phi\sigma^r_\phi=-2\eta(g_{rr}/g_{\phi\phi})[u^r(L-L_0)/(2\nu)]^2$. If we assume viscosity in the flow is zero then the equation (\ref{ec.eq}) becomes the canonical relativistic Bernoulli parameter and can be written as $B=-hu_t$ \citep{kc17}.

Now, we have simplified the equations (\ref{rNS.eq} and \ref{reg.eq}) with the help of the equations (\ref{pNS.eq}-\ref{mdot.eq} and \ref{eos.eq}) then we get the expression of velocity gradient
\begin{equation}
\frac{dv}{dr}=\frac{\frac{\as^2N}{N+1/2}\left[H_r+\frac{2}{r}-\frac{g_{rr,r}}{2g_{rr}} \right]+\frac{2\nu\sigma^{r\phi}\sigma_{r\phi}}{h(N+1/2)u^r}-{\cal F}}{\gamma_v^2\left[v-\frac{\as^2N}{v(N+1/2)}\right]}=\frac{\cal N}{\cal D},
\label{dv.eq}
\end{equation}
and the temperature gradient
\begin{equation}
\frac{d\Theta}{dr}=-\left[\Theta\left(\frac{u^r_{,r}}{u^r}+\frac{2}{r}+H_r\right)+\frac{\tilde{\tau}\nu}{u^r}2\sigma^{r\phi}\sigma_{r\phi}\right]/[N+1/2],
\label{dth.eq}
\end{equation}
where $H_r=0.5/r+[2r(r^2+a^2)-\Delta^{'}a^2]/[(r^2+a^2)^2-2\Delta a^2]-[2r(r^2+a^2)+\Delta^{'}a^2]/[(r^2+a^2)^2+2\Delta a^2]$.
Using the equation (\ref{pNS.eq}) into the equation (\ref{srp1.eq}) then we get the expression of angular momentum gradient
\begin{equation}
\frac{dl}{dr}=-\frac{g_{rr}}{\gamma_v^2}\left[\frac{u^r}{\nu}(L-L_0)+\frac{1}{3}u^ru_\phi u^r_{,r}+\tilde{\sigma}\right],
\label{dl.eq}
\end{equation}
 where $u^r=\gamma_v v/\sqrt{g_{rr}}$, $\gamma_v=1/\sqrt{1-v^2}$ is bulk Lorentz factor, and $v$ is bulk three-velocity of the fluid.
\section{Numerical Method}\label{sec:nmethod}
\begin{figure}[htbp]
\begin{center}
 \includegraphics[angle=0, width=0.50\textwidth]{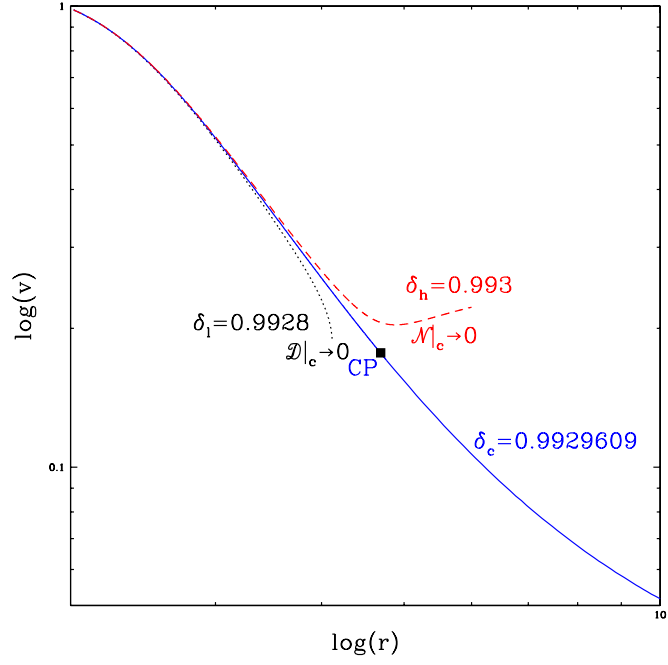}
\caption{Variation of $v$ with $r$ for three values of iteration parameter $\delta=0.9928$ (dotted black), $0.9929609$ (solid blue), and $0.993$ (dashed red) are plotted. A symbol $\blacksquare$ represents the CP location $r_c=4.76823$. The disk parameters $E=1.001, L_{in}=3.3, \xi=1, \alpha=0.001~\&~a=0.1$ are used.}
\label{Fig1R}
\end{center}
\end{figure}

\begin{figure}[htbp]
\begin{center}
 \includegraphics[angle=0, width=0.45\textwidth]{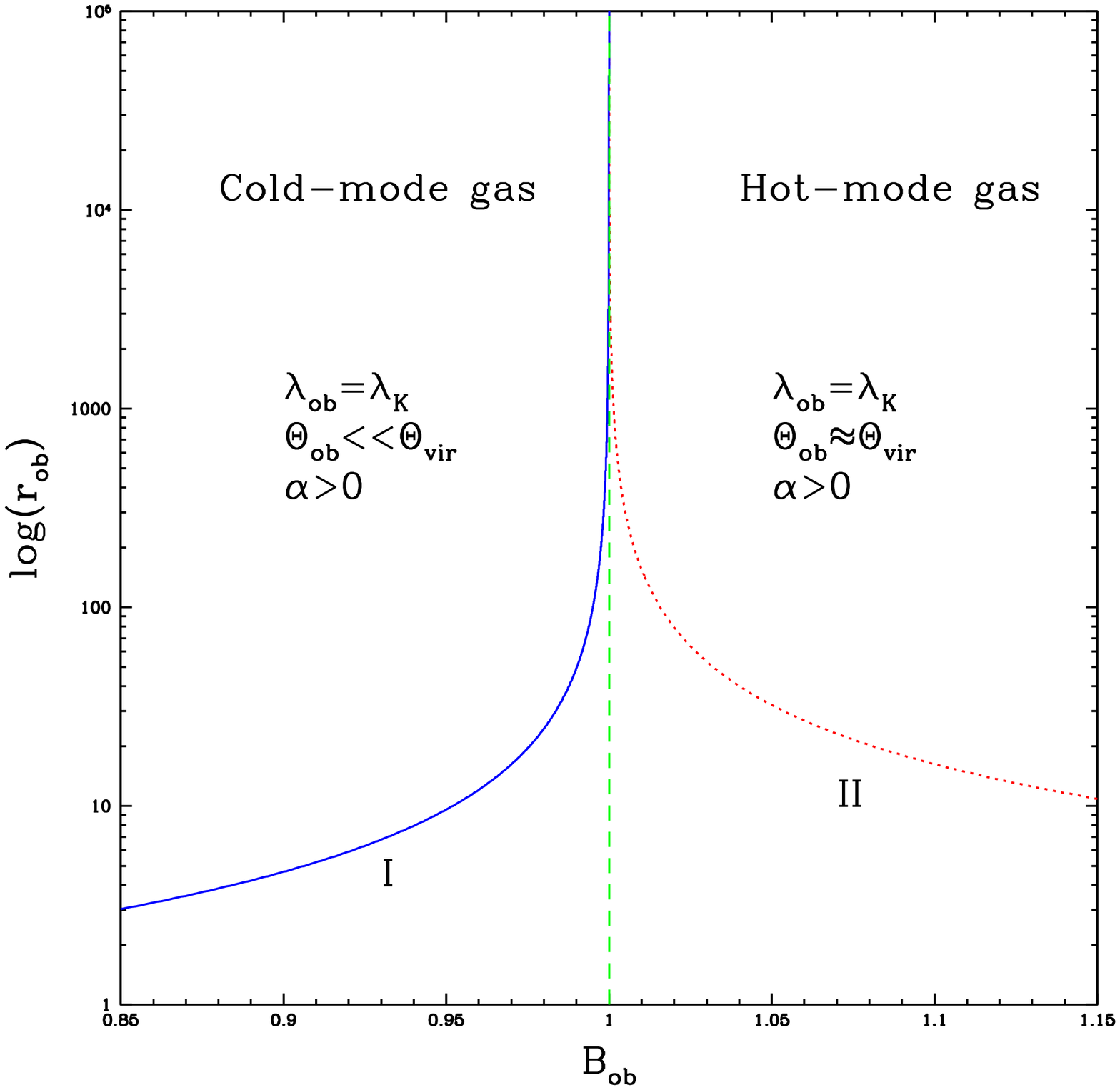}
 \includegraphics[angle=0, width=0.45\textwidth]{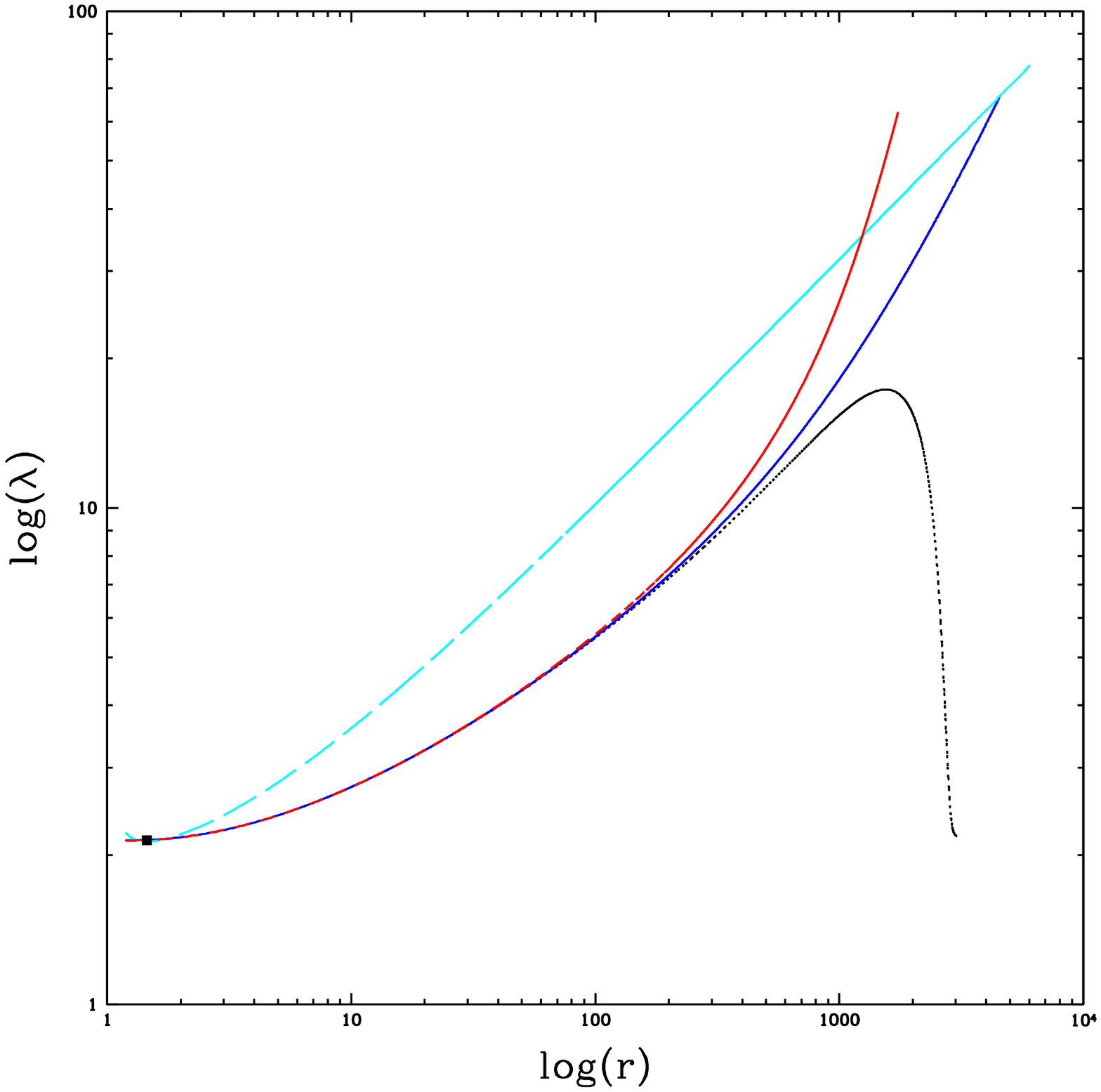}
\caption{We represented the variations of $\rout$ with $\be$ in the left panel and  the $\lambda$ distribution with the radial distance in the right panel. Here $\rout$ is the possible outer AB location with $\lmdob=\lmk$ of the solutions with corresponding local energy ($\be$). 
In the right panel, each curve is plotted for the same parameters $E=1.025, \xi=1.0, \alpha=0.001$ and $a=0.99$ with the different $L_{in}|_{l}=2.1809$ (dotted, black curve), $L_{in}|_{h}=2.1891$ (dashed, red curve), and $L_{in}|_{c}=2.18909323$ (solid, blue curve). A long-dashed cyan curve represents the Keplerian AM distribution in the right panel.}
\label{Fig2R}
\end{center}
\end{figure}

We have solved three differential equations (\ref{dv.eq}-\ref{dl.eq}), simultaneously by the 4th order Runge-Kutta numerical method. Mathematically, these differential equations have infinite number of solutions for an infinite number of accretion boundary conditions (ABCs) but they may or may not be physical solutions depending on the nature of an object. As we knew that an accretion solution must be transonic in nature \citep{B52}.
Thus we have developed three iteration schemes to get the physical advective transonic accretion solutions around the BHs as follows:\\
I. {\it Iteration to get critical point (CP)}:  The mathematical expression of critical point conditions (CPCs) are obtained by the equation (\ref{dv.eq}) via putting $dv/dr={\cal N}/{\cal D}=0/0$ then 
\begin{eqnarray}
{\cal D}|_c=0 \Rightarrow v_c=\sqrt{\frac{2}{\Gamma_c+1}} {\as}_c, \label{dc.eq}\\
{\cal N}|_c=0 \Rightarrow \frac{2{\as^2}_c}{\Gamma_c+1}\left[H_{r}|_c+\frac{2}{r_c}-\frac{g_{rr,r}|_c}{2g_{rr}|_c}\right]+\frac{g_{rr}|_c}{g_{\phi\phi}|_c}\frac{u^r_c(L_c-L_0)^2}{h_c(2N_c+1)\nu_c}-{\cal F}|_c=0
\label{nc.eq}
\end{eqnarray}
where suffix `$c$' represents flow quatity at the CP location ($r_c$).
Since $E$ is constant throughout the flow therefore the equation (\ref{ec.eq}) can be written as
\begin{equation}
E=E_c=h_c\gamma_{v_c}\sqrt{r\Delta}~exp\left[{\cal F}_p|_c+\frac{g_{rr}|_c}{g_{\phi\phi}|_c}\frac{u^r_c(L_c-L_0)^2}{h_c\nu_c}\right]dr_c,
\label{ecc.eq}
\end{equation}
where, $d\rc=\rc-\rh$ or $\rc-\rout$. There are four unknown variables ($\rc, v_c, \Theta_c$ and $L_c$) and three equations (\ref{dc.eq}, \ref{nc.eq} and \ref{ecc.eq}) at the CP. Thus, this is a problem of unknown variables and we can not directly solve the CP as done in many previous studies with some assumptions or simple form of viscosity in the relativistic/non-relativistic regime \citep{acg96,lgy99,cd07,ddmn19}. Here we have used iteration technique to get the CP location for given five model parameters, which are the two integration parameters ($E$, $L_0$ or $L_{in}$), two fluid parameters ($\xi, \alpha$) and a BH spin parameter ($a$), where $L_{in}$ is very close to the BH horizon, say, $r_{in}=\rh+0.001, \rh=\rg+\sqrt{\rg^2-a^2}$ and $\rg=1$. We also have used this technique in our previous studies \citep{kc14,kc15,ck16}. In this iteration method, we have started the integration of the differential equations from very close to the BH, say $r_{in}$ to outward. Doing so, we need to estimate the three flow variables, and denoted as $v_{in}, \Theta_{in}$ very close to the BH horizon and $L_{0}$ at $\rh$ with supplying five parameters ($E, L_{in}, \xi, a$ and $\alpha$). Here it is convenient to supply $L_{in}$ at $\rin$ instead of $L_0$ at $r_h$  and sophistically calculated $L_0$ in below an equation (\ref{lin.eq}). For the starting of the iteration, We assumed that the bulk velocity very close to the BH is a like free-fall as
\begin{equation}
v_{in}=\delta\sqrt{\frac{r_h}{r_{in}}},
\label{vin.eq}
\end{equation}
where $\delta<1$ is an iteration parameter with some initial guess value and actual value of it can be obtained after many iterations when CPCs are satisfied.
Following \cite{kc15} and assumed $E=B$ since $\tau^r_\phi\rightarrow 0$ from the equation (\ref{pNS.eq}) as $L\rightarrow L_0$ when $r\rightarrow r_h$ then from the equation (\ref{ec.eq}), we get as 
\begin{equation}
{\cal F}+Q^+=\frac{1}{u_t}\frac{du_t}{dr}-\frac{1}{\gamma_v}\frac{d\gamma_v}{dr},  ~~\mbox{at}~~r=\rin\rightarrow\rh
\label{eh.eq}
\end{equation}
where, $Q^+=q^+/(\rho u^rh)$ and $u_t=-wl-\sqrt{-1/g^{tt}}\sqrt{\gamma_v^2+l^2/g_{\phi\phi}}$. Now simplifying above equation with using the equations (\ref{dv.eq}, \ref{dl.eq}) then wrote in the power of $(L_{in}-L_0)$ as
\begin{equation}
A_2(L_{in}-L_0)^2+A_1(L_{in}-L_0)+A_0=0,
\label{lin.eq}
\end{equation}
where, $A_2=u_r[1-u_{tlv}/\{{\cal D}(2N+1)\}]/(h\nu g_{\phi\phi})$, $u_{tlv}=[\{u_{tl}l/3-u_{tv}\gamma_v^2\}/u_t-1]v\gamma_v^2$, $u_{tl}=w+u_{tv}l/g_{\phi\phi}$, $u_{tv}=\sqrt{-1/g^{tt}}\sqrt{1/\{\gamma_v^2+l^2/g_{\phi\phi}\}}$, $A_1=u_{tl}u_r/[u_t\nu\gamma_v^2]$, and $A_0={\cal F}-u_{tlc}/u_t-u_{tlv}[2\as^2\{2/r-0.5g_{rr,r}/g_{rr}+H_r\}/\{\Gamma+1\}-{\cal F}]/{\cal D}$, $u_{tlc}=u_{tc}+g_{rr}u_{tl}[\tilde{\sigma}-lu^ru^rg_{rr,r}/(6g_{rr})]/\gamma_v^2$, and $u_{tc}=-lw_{,r}+u_{tv}[(\gamma_v^2+l^2/g_{\phi\phi})g^{tt}_{,r}/g^{tt}+g_{\phi\phi,r}/g^2_{\phi\phi}]/2$.
Suppose $L_{in}-L_0=\zeta$, thus obtained $L_0=L_{in}-\zeta$ and considered $L_0<L_{in}$ because outward transportation of the AM in the accretion due to the viscosity. 
At $r_{in}$ the $\Theta_{in}$ is obtained from $E=B_{in}=-h_{in}u_t|_{in}$ as
\begin{equation}
\frac{f|_{in}+2\Theta_{in}}{\tilde{\tau}}-\frac{1}{\gamma_{v_{in}}}\sqrt{-g^{tt}(B_{in}-wL_{in})^2-\frac{L_{in}^2}{g_{\phi\phi}}}=0.
\label{thin.eq}
\end{equation}
Now we have solved equations (\ref{vin.eq}, \ref{lin.eq} and \ref{thin.eq}) together and obtained the local flow variables, $v_{in}, \Theta_{in}$ and $L_0$ with a given value of the $\delta$ and disk parameters, $E, L_{in}, \xi, a, \alpha$.  After obtaining the value of $L_0$ and other local variables at $r_{in}$ then we can integrate the all three equations (\ref{dv.eq}, \ref{dth.eq}, \ref{dl.eq}) outward from $r_{in}$ with checking the CPCs (\ref{dc.eq} \ref{nc.eq}). If the CPCs are not satisfied then we changed the value of $\delta$ in some rational way depending upon variation tendency of the local flow variables, which is illustrated in a Figure (\ref{Fig1R}).
As we know that the CP can be obtained when both the CPCs must be satisfied, simultaneously, which we got for $\delta_c=0.9929609 $ (blue curve) say, $correct ~value$ for this particular case, and other values of it can only satisfy one CPC either ${\cal D}\rightarrow 0$ ($\delta_l=0.9928, lower ~value$, black curve) or ${\cal N}\rightarrow 0$ ($\delta_h=0.993, higher ~value$, red curve).  With the correct value of $\delta_c$, we got the $r_c$ and corresponding $v_c, \Theta_c, L_c$. Now, we can calculate the slops with using these flow variables at the CP by L'hospital rule and the equation (\ref{dv.eq}) can be written as $(dv/dr)|_c=(d{\cal N}/dr)|_c/(d{\cal D}/dr)|_c$ and after some simplifications with using other two equations then we get  a quadratic equation
\begin{equation}
{\cal A}_2\left(\frac{dv}{dr}|_c\right)^2+{\cal A}_1\frac{dv}{dr}|_c+{\cal A}_0=0,
\label{dvc.eq}
\end{equation}
where ${\cal A}'s\equiv f(r_c, v_c, \Theta_c, L_c)$  and we have avoided to write these tedious expressions just to save time and space. 
This equation has two roots, one with negative sign gives the accretion and other positive sign gives the wind like solutions. In the present study, the accretion flow is only investigated. After the obtaining of $(dv/dr)|_c$, and the other derivatives from the equations (\ref{dth.eq}, \ref{dl.eq})  are also obtained at the CP. 
Now we can further integrate the equations from the $r_c$ in the 
outward direction from the center of the BH. Thus we got the complete transonic accretion solution as represented by middle curve of the Figure (\ref{Fig1R}).  
Only finding of the inner CP is not ensure the physical global solution, since accretion solution may have 2-CPs or 3-CPs solution and the detail classification is presented in a parameter space as shown in a Figure (\ref{Fig4R}). Therefore we have made two more iteration schemes in order to get required outer ABCs and all possible kind of accretion solutions around the BHs. 
\\
II. {\it Iteration to get ADAF solution}: This iteration can executed only after getting the inner CP (1st iteration). For the investigations of the ADAF solutions, first, we made the rough classification of the parameter space that can give us the guess values of the parameters, which can have the ADAF solutions. 
We believed that the maximum AM and temperature of the accreting gas can be the Keplerian ($\lmk$) and a virial temperature ($\thvir$) at the outer AB, respectively, and both are varying with the radial distance. So we can define the ABCs of the flow as $\lmdob\le\lmk$ and $\Thob\le\thvir$ with $v\rightarrow 0$ at an outer AB location ($\rout$). The $\Thob$ can be different for the accreting gas at the $\rout$. So we changed the $\Theta$ of the gas and calculated the local energy of the gas, $\be=-hu_t|_{\rout}$ with the variation of the radius, and assumed $\lmdob=\lmk$ for the maximum $\rout$ in the viscous flow.
Thus these conditions can give us broadly two type of the ABCs with their the AB locations, $\rout$ and the local energies, $\be$.
The variations of the $\rout$ is divided $\be$ into two regions with the help of the $\Thob$ as represented in a left panel of a Figure \ref{Fig2R} \citep[For the first time, the same kind of figure has been generated with pseudo-Newtonian geometry by][]{kg19}. The interesting thing of this division is that the $\be<1$ can have one or two CP solutions so there is a possibility to give the ADAF solutions \citep[1-CP][]{ny94}, and $\be>1$ has one or three CP, which has possibilities to provide Bondi-type \citep[outer CP][]{B52}, shock type solutions \citep[3-CPs][]{c89}, and smooth solutions or  ADAF-thick \citep[inner CP][]{lgy99}. So, this division of the $B$ is also simplified the problem of choosing parameters of the flow, when we were hunting for the various kind of the accretion solutions.

In this iteration scheme, we have investigated  the outer ABCs for the ADAF solutions with following conditions of \cite{ny94,kg19}, which are $\lambda_{ob}=\lambda_K, v\rightarrow 0~\mbox{and}~
\Theta_{ob}<<\Theta_{vir}$ (gives $\be<1$) at the $\rout$ and the scheme is illustrated in the right panel of the Figure (\ref{Fig2R}). Here $L_{in}$ is the our iteration parameter and changed it to till satisfied the ABCs. 
In the right panel of the Figure (\ref{Fig2R}), we have represented three iterations with different $L_{in}$. For $L_{in}|_{h}$, the $\lambda$ distribution (red curve) is high and exceeds the $\lmk$ distribution in the outer AB. Moreover, $B$ is increasing outward after minima and $\be>1$ with $\Theta>\Theta_{vir}$ at $\lambda=\lambda_K$. For $L_{in}|_{l}$, $\lambda$ is low (black curve), $B$ is less than $1$ at $\rout$ but $\lambda<<\lambda_K$ and behaviour of $v$ is also unusual (increasing outward). For the correct ADAF solution, all the outer ABCs must be satisfied, simultaneously. Which is found by the iteration with some rational way variation of the $L_{in}$, and for this particular case, the correct value of it is $L_{in}|_{c}=2.18909323$ (blue curve) and corresponding $\be=0.999894$ at $\rout\approx4553$. 
\\
III. {\it Iterartion to find shock solution}: In this part of iteration scheme, we have investigated the multiple critical point (MCP) shock solutions and the outer ABCs are  $\lambda_{ob}=\lambda_K, v\rightarrow 0~\mbox{and}~0.1\times\Theta_{vir}\lsim\Theta_{ob}<\Theta_{vir}$ (gives $\be>1$) at the $r_{ob}$. Doing so, we need to find the inner and outer CP for a particular disk parameters. Here we assumed the non-dissipative shocks in the flow, which means there is no mass-loss and radiative loss. Thus the adiabatic relativistic shock conditions \citep{t48} are the conserved form of mass flux [$\dot{M}$], momentum flux [$(e+p)u^2+W$], angular momentum flux [$\dot{J}=\dot{M}L$] and energy flux [$\dot{E}=\dot{M}E$] across the shock location can be written as \citep{kc15,ck16},
\begin{eqnarray}
[\dot{M}]=0 \Rightarrow \rho_+u_+H_+=\rho_-u_-H_-, \label{mf.eq}\\ ~
[\Sigma hu^2+W]=0\Rightarrow h_-u_-^2-{\cal K}_mu_-+\frac{2\Theta_-}{\tilde{\tau}}=0, \label{momf.eq}\\ ~
[\dot{J}]=0\Rightarrow L_-=L_++2\sigma^r_{\phi}|_+\left(\frac{\nu_+}{u_+}-\frac{\nu_-}{u_-}\right), \label{jf.eq}~\mbox{and}~\\ ~
[\dot{E}]=0\Rightarrow {\cal K}_{\cal E}-h_-\gamma_{v_-}exp(X_{f_-})=0,
\label{ef.eq}
\end{eqnarray}
respectively. The equation (\ref{mf.eq}) is used to simplify the other three equations, and suffix `$+$' and `$-$' represents the post-shock and pre-shock flow variables, respectively. Here ${\cal K}_m=[h_+u_+^2+2\Theta_+/\tilde{\tau}]/u_+$, ${\cal K}_{\cal E}=h_+\gamma_{v_+}{\rm exp}(X_{f_+})$, $u=\gamma_vv$, $X_{f_-}={\cal F}_{\lambda}\int{\cal F}_{p+}dr+f_q\int Q^+_+dr$, ${\cal F_{\lambda}}={\cal F}_{p-}/{\cal F}_{p+}$, $f_q=Q^+_-/Q^+_+$, $2\sigma^r_{\phi}|_+=-u_+(L_+-L_0)/\nu_+$, and the remaining terms are already mentioned in the current section and previous section. 

\begin{figure}
\begin{center}
 \includegraphics[angle=0, width=0.50\textwidth]{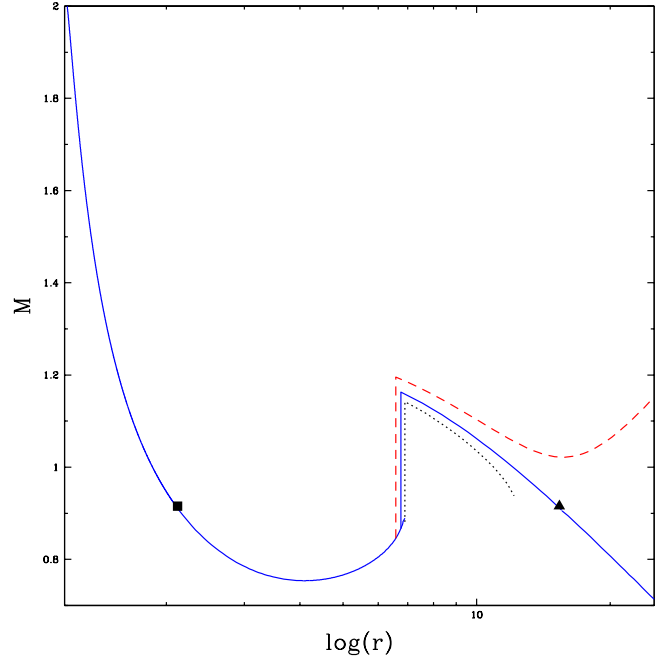}
 \caption{Variation of the Mach number ($M$) with the radial distances ($r$) is presented. The solid curve represents the correct shock solution, which is passed through the inner CP ($\blacksquare$ point) and outer CP ($\blacktriangle$ point).  The vertical lines represent the tentative shock locations, $\rs|_{low}=6.57$, $\rs|_{correct}=6.75374456$, and  $\rs|_{high}=6.82$. The corresponding shock locations represented the super/sub-sonic branches of the flow but only correct one passed through the outer CP.
 \label{Fig3R}}
 \end{center}
\end{figure}
Now we have supplied the flow parameters, which correspond to the region II of the left panel of the Figure (\ref{Fig2R}). A first step is to investigate the inner CP by the 1st iteration scheme and the second step is to check the closed topology solution ($\alpha-$type) with $B>1$ when we integrate outward from the inner CP. If we get $\alpha-$ type solution then go ahead for the calculation of the super-sonic flow variables at the presumed shock location with the help of the shock equations (\ref{momf.eq}-\ref{ef.eq}), which is illustrated with the Mach number ($M=v/a_s$) variations in the Figure \ref{Fig3R}. 
After over of 1st iteration scheme, we started to integrate the fluid equations outward from the inner CP and checked the variation of $M$ with the radial distance. When we got $dM/dr>0$ then we began to calculate the flow variables of the super-sonic branch with the help of the shock equations (\ref{momf.eq}-\ref{ef.eq}) at the presumed shock location and checked the CPCs (\ref{dc.eq} and \ref{nc.eq}) for the outer CP. Here, the iteration parameter is radial distance with presuming as the shock location as represented in the Figure \ref{Fig3R}. At the presumed shock location we calculated the super sonic flow variables with help of the shock equations then we again integrated outward and the super sonic branches are shown in the Figure \ref{Fig3R} with three different presumed shock locations, $r_s|_{low}=6.57$ (dashed red curve), $r_s|_{high}=6.82$ (dotted black curve) and $r_s|_{correct}=6.75374456$ (solid blue curve), which passes smoothly from the outer CP. The super-sonic branch corresponding to $r_s|_{low}$ is always having the super-sonic nature with ${\cal N}|_c\rightarrow 0$ and the branch with $r_s|_{high}$ is truncated with ${\cal D}|_c\rightarrow 0$ at some distance. So the situations are the same as in the 1st iteration scheme so in order to get the CP location, both CPCs must be satisfied, simultaneously. Doing so, we again made the iteration scheme with the iteration parameter that can investigate the shock location and outer CP with satisfying CPCs. After getting outer CP and corresponding flow variables, we again integrated outward from the outer CP so we got full accretion solution with shock as represented with solid line (blue curve) in Figure \ref{Fig3R}.

\section{Results}\label{sec:result}
We used five disk flow parameters ( ${E}, L_{in}, a, \xi$ and $\alpha$) to investigate the transonic accretion solutions on the equatorial plane. 
The value of parameters $a=0.99, \xi=1$ and $\alpha=0.001$ are kept same and we changed the other two integration parameters in the rest of the study. 
Doing so, we have investigated all possible transonic advective accretion solutions, which are presented and discussed in the following sub-sections with their possible physical perspectives. 
Since we knew that the possible nature of the advective solutions can not be much changed with changing of the $\alpha$ upto certain values \citep{kc13,kc14,ck16} or with the full range of $a$ and $\xi$ \citep{kc17}, except for $\xi\sim0$ case \citep{kscc13,kc17}. Therefore, we have not changed these parameters in this study.
\subsection{Parameter space in the $\be-L_0$ plane}\label{subsec:ps}
There are many studies have done in order to investigate the region of parameter space (PS) by exploring all the possible transonic accretion solutions and represented them on one plane of energy and AM space \citep[][references therein]{kc17}. But all the studies have missed the ADAF solution \citep{ny94} to locate in the PS. Therefore for the first time, we are going to present the PS with including the location of the ADAF solutions. We will also discuss the details about the possible physical applications and their possible reasons for the formation of these kind of solutions in the following subsections.

\begin{figure}[ht!]
\plotone{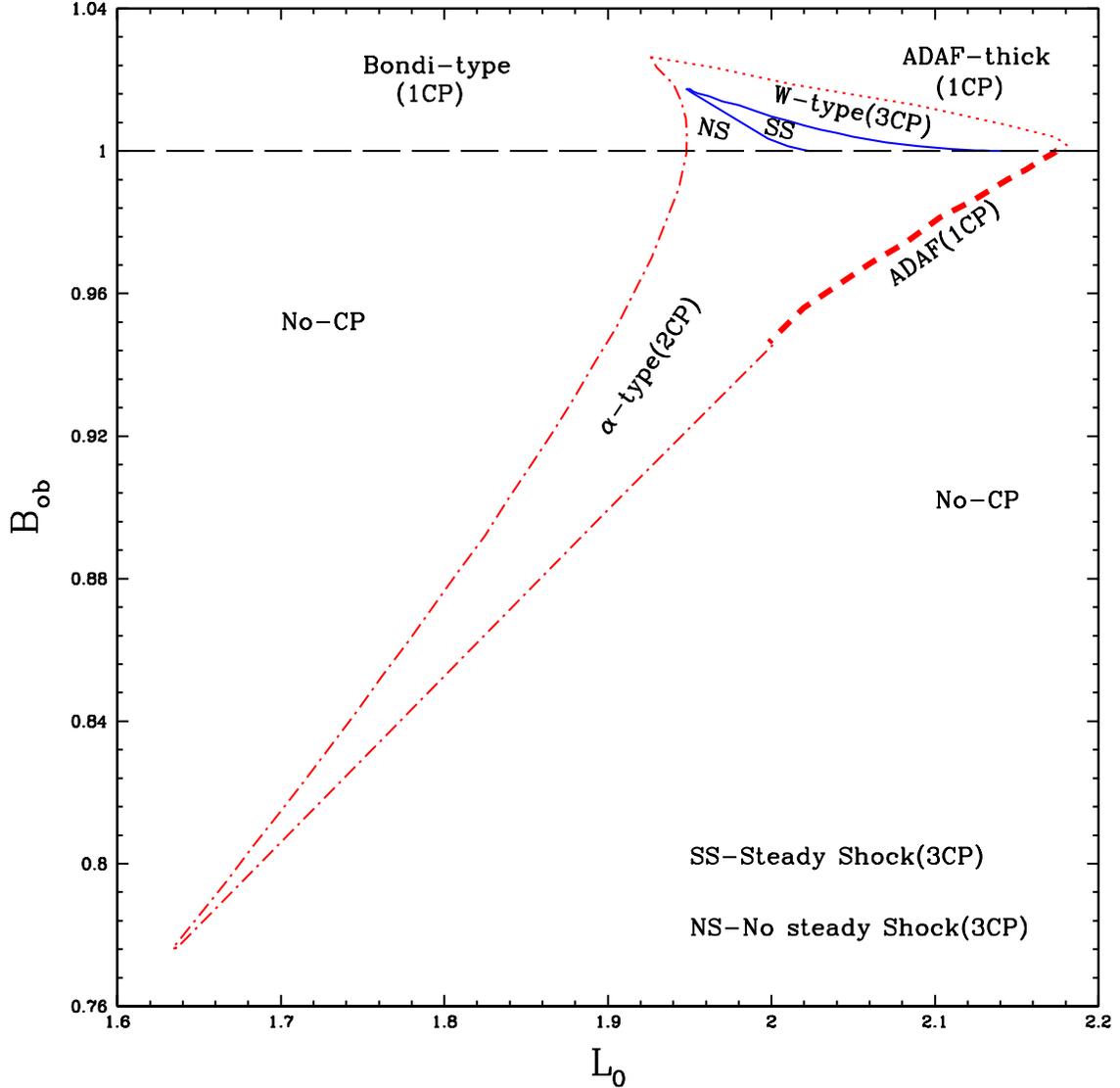}
\caption{The division of the PS in $\be-L_0$ are represented with all possible transonic solutions of the GRHD flow. The PS is broadly divided into eight regions on the basis of type of the advective solutions with including NS \citep[oscillatory/unstable shocks][]{lckhr16} region and No-CP region. 
\label{Fig4R}}
\end{figure} 
The accreting gas at the AB can be the Keplerian or sub-Keplerain AM with \hm or \cm or both \citep[for example, a cold clumpy cloud surrounded by hot dilute gas][]{byz20}. 
So, for the detail understanding of their respective solutions, we have investigated the PS in $\be-L_0$ and divided into many regions in the Figure \ref{Fig4R}. 
The PS is represented with the $\be$ because our main intention is to relate the kind of the advective solutions with their nature of the gas at the outer AB locations 
as roughly presented in the left panel of Figure \ref{Fig2R}.
Here the PS is broadly divided into eight regions on the basis of the typical behavior of the solutions and number of CPs, like, 3CPs, 2CPs, 1CP and no CP regions. The 3CPs region is again divided into steady shock (SS), no-shock (NS) or oscillatory shock and $W-$type solutions with $\be>1$. The $W-$type name is adopted from the previous studies \cite[][and references therein]{kc17} because of the multivalued nature and variation of the Mach number. 
The 1CP region with $\be>1$ is divided into the Bondi-like flow with the outer CP, and the smooth or ADAF-thick flow with the inner CP. The 1CP with $\be<1$ gives the ADAF (thick-dashed line), and the 2CP region with $\be<1$ gives the $\alpha$ type solutions. Interestingly, the solutions in the 2CP region with $\be<1$ can be only formed when $\lmdob<\lmk$ and discussed more in the sub-section \ref{subsec:ups}.
From this PS plot, we can understand the possibility of the occurrence of transonic accretion solutions, which can determine the inner structure/geometry of the accretion flow.  
The SP regions $\be<1$ and $\be>1$ corresponds to the outer AB locations of the region ${\rm I}$ and ${\rm II}$ in the left panel of Figure \ref{Fig2R}, respectively. 

The PS region of the standing shock is occurred in the middle range of the bulk AM with the energy ($>1$) in the Figure \ref{Fig4R}. The shock can not be formed in the full MCPs region (3CPs region denoted with dotted and part of the dash-dotted curve with $\be\ge1$).
Since the formation of the shock needs the sufficient AM distribution and long-range flow in order to become super-sonic at the sufficiently large distance (details in the next sub-section). Therefore the energy range is low compared to the other kinds of the solutions with $\be\ge1$ and the low $\be$ with the hot flow has large AB, which is also cleared from the region II of the left panel of Figure \ref{Fig2R}.
Moreover, the standing shocks are very sensitive to the viscosity, radiative cooling, mass-loss and composition of the fluid when compared with the other kind of flows \citep{kc14,kc17}. Since the shock flow needs sufficient $\lambda$ that can form temporary centrifugal boundary layer in the flow against to the gravity \citep{c89} and make the preceding flow enough compressed to create the shock \citep{fkr02} therefore the shocks can not possibly be formed at the very high viscosity \citep[$\alpha\gsim0.3$,][]{kc14} and unlike to the other advective solutions \citep{nkh97,kc13,kc14,ck16}.

The PS region for the ADAF flow \citep{ny94} is narrow for a one value of the $\alpha$ but the $\be$ range is large between $\sim0.94$ to $1$ as represented with the thick-dashed line in the Figure \ref{Fig4R}. 
The origin of this kind of solutions seem very specific with very unique outer ABCs particularly, $H/r\ll1$ at the $\rout$ (see solution of a Figure \ref{Fig7R}). Therefore with this one more reason, we believed that they are generated from the gas of the inner KD means, the KD is like a gas source for the ADAFs. 
This has been discussed in the many studies for the unification theory of the hybrid disk \citep{ny94,h96,mk00,gpkc03,mm94,gl00,mlm00,llg04}, and recently, \cite{kg19} have also been shown that the local energy of the KD at the inner boundary is matched with the local energy of the ADAFs at the outer AB. The whole PS also gives the information about the same $\be$ or $\rout$ can give the different kinds of physical solutions with $\be>1$ (Figure \ref{Fig5R}), except $\be<1$ which has only a ADAF solution with an unique outer AB location (see a panel c of a Figure \ref{Fig8R}).

Moreover, the PS region of the Bondi-like flows $0\le L_0\lsim1.9$ (for used particular viscosity parameter in this study) is largest from the other kind of solutions, and placed on the low AM side in the PS. It can also exist for large range of the mass accretion rate and viscosity \citep{kc14}. Unlike, the PS region of the other kinds of flows shrink very fast with increasing the viscosity parameter or mass accretion rate, specially MCPs regions \citep{kc14}. 
The Bondi-like solutions are passed only through the outer CP and has lowest AM distribution. So they are fastest and highly radiative inefficient \citep{st83}. 
Thus, we can conjecture  that the Bondi-like flows maybe good for the BH growth scenario since they are slowest rotator (quasi-spherical), fast inward motion and highly radiative inefficient flows, so which can have very low possibility of the wind production as also seen in the simulation \citep{lckhr16}.

For the comparison, we have listed some of extreme properties of the different solutions in Table \ref{tab:tab1}, which can help in understanding of their physical feasibilities with their sizes, temperatures etc. Here we defined the monotonic and multivalued nature of the flow on the basis of the variation of Mach number ($M$) with the radial distances of the solutions as represented in the previous studies \citep{kc14,kc17}. The multivalued nature of a solution means the $M$ is found same at more than one radial distance.
The reason for the nature of multivalued solutions are different with type of solutions, such as, the shock solutions have found multivalued mostly close to the BH when shock occurred, $W$-type found multivalued around the outer CP location due to tendency of the flow can passed through the outer CP, and $\alpha-$type with $\be<1$ found multivalued due to second/middle CP.
 The monotonic nature of the solutions are called the smooth solutions, like, the ADAF and the ADAF-thick. The ADAF-thick solutions \citep{lgy99} are geometrically more thicker than the ADAF \citep{ny94}.
 In the table, we represented the minimum sizes of the kind of solutions since the maximum size of the accretion solutions can be very-2 large when $\be\sim1$.
 Moreover, the transportation of the AM also depend on the location of the outer AB, which can also decide the possible nature of the flow in the disk, for instance, the shock can not form with $\rout<10^3$ (details in the following subsection). 
The variation of values of the tabulated quantities like $\rout|_{min}, \Thmx$ are strongly depend on the $\be$ with achieving the Keplerian AM at the $\rout$ (Figures \ref{Fig2R} and \ref{Fig8R}c), except the $\alpha-$type solutions which are entirely sub-Keplerian flow. Other quantities like $R, \rs$ are also strongly dependent on the $\be$ and $L_0$ at the horizon (Figure \ref{Fig8R}a). Here $R=\Sigma_+/\Sigma_-=u_-/u_+$ is the compression ratio at the shock location and $\Sigma=2\rho H$ is the surface density of the flow. 
Here the $R$ tells the strength of the shock in the flow, if it is higher then the shock is stronger and the temperature is raised more. 
Since, the density and temperature is sharply raised at the $\rs$ in the post-shock flow, so this region can act like a corona in the disk. Therefore, the $R$ can also quantify the effectiveness of the inverse-Comptonization of the radiation \citep{ct95,mc10,kcm14}. 
\begin{table}
\caption{We represented the some properties of the hot advective solutions for the $a=0.99$ and $\xi=1$. Here $\rout|_{min}$ is a minimum size of a particular type of solution which commonly represented in the left panel of Figure \ref{Fig2R}. $\Thmx$ is a maximum temperature obtained for a particular type of solutions with their corresponding energy range $\be$.
}
\begin{tabular}{c c c c c c}
\hline \hline
Type of solution&Local energy&Min. disk size&Max. temperature&Compression&Shock location\\
--&at OBL $B_{\rm ob}$ & $\rout|_{min}$ & $\Thmx$ & ratio $R$&$\rs$ \\
\hline
ADAFs (monotonic) & $\le1$ & $\sim6$  & $50\lsim\Theta_{max}\lsim100$ & NO & NO \\
\hline
$\alpha$-type (multivalued) & $\le1$ & $\sim2$  & $20\lsim\Theta_{max}\lsim100$ & NO & NO \\
\hline
Shocked (multivalued) & $\ge1$& $\sim5\times10^3$ & $\sim180$ & $1<R\lsim5$ &$4\lsim r_s\lsim4000$ \\
\hline
W-type (multivalued) & $\ge1$ & $\sim10^3$ & $100\lsim\Theta_{max}\lsim180$ & NO & NO \\
\hline
Smooth (monotonic) & $\ge1$ & $\sim3$ & $\gsim100$ & NO & NO\\
\hline
Bondi-type (outer CP) & $\ge1$ & $\sim10^3$ & $\gsim100$ & NO & NO\\
\hline
\end{tabular}
\label{tab:tab1}
\end{table}

\subsection{The advective disk solutions and structures}\label{subsec:hybriddisk}
\begin{figure}[ht]
\plotone{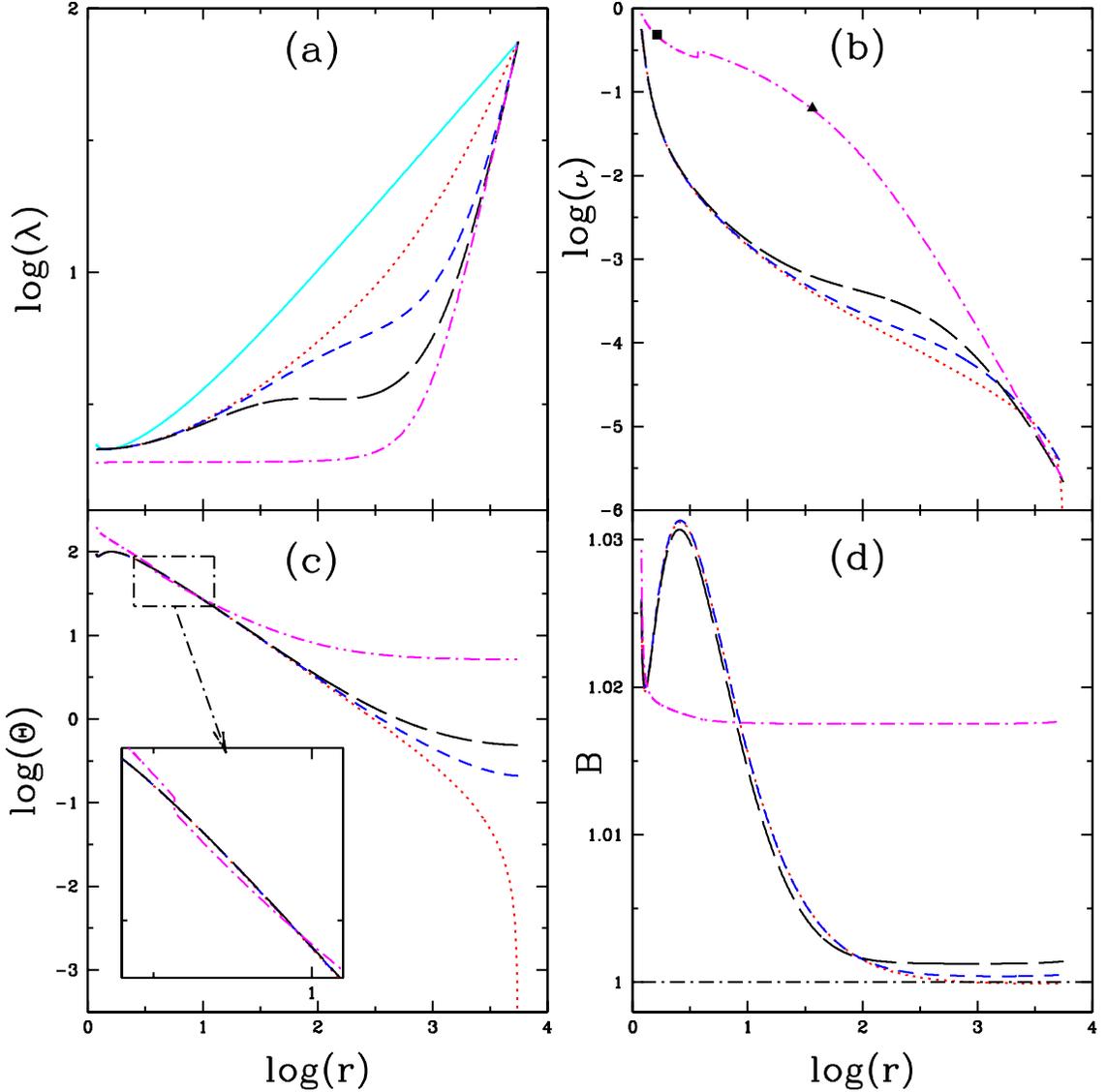}
\caption{Variations of flow variable $\lambda$ (panel a), $v$ (panel b), $\Theta$ (panel c) and $B$ (panel d) with the $r$ are presented  with keeping same outer AB location $\rout\sim5500$ and different $\be=0.99991$ (dotted), $1.00052$ (dashed), $1.00141$ (long-dashed), and $1.01768$ (dash-dotted). 
 The dotted (red), dashed (blue), long-dashed (black), and dash-dotted (magenta) curves are representing four kinds of accretion solution, ADAF, smooth, $W$-type and shocked flow, respectively. A solid curve (cyan color) represents Keplerian AM distribution in the panel (a).
\label{Fig5R}}
\end{figure}
We have represented the typical transonic advective solutions with keeping $\rout\sim5500$ is fixed in the Figure \ref{Fig5R} with the different disk parameters of the four different regions of the PS in Figure \ref{Fig4R}. The values of $\be$ of the each solution is mentioned in the caption and the $\be<1$ represents the \cm gas, and $\be>1$ represents the \hm gas at the AB.
The dotted, dashed, long-dashed, and dash-dotted curves are representing the ADAF, smooth (or ADAF-thick), $W$-type, and shock solutions, respectively. The ADAF and ADAF-thick solutions have represented for the single inner CP. The $W$-type and shock solutions have represented from the MCPs region. The shock solution is passaged through the two-CPs (inner and outer CP) and the $W-$type solution is passaged through only the inner CP. 
In the panel (a) of the Figure \ref{Fig5R}, the AM distribution of the flows are plotted. The ADAF (dotted curve)  and shocked (dashed-dotted curve) flows have the highest and lowest AM distribution, respectively, and the $\lambda$ is transported by the viscosity of the flow. 
Although, the dependence of viscous stress tensor ($\tau^r_\phi=-2\eta\sigma^r_\phi$) looks very complicated as from the equation (\ref{srp1.eq}), but  $\tau^r_\phi\propto\Theta$ seems to be more effective around the AB with the higher $\Theta$ gas. 
Since the shock flow has higher $\Theta$ at the $\rout$ (panel c) therefore, the AM distribution is decreased very fast around the AB. The resulting is that the flow velocity is raised very fast (dash-dotted curve, panel b) and become the super-sonic far away from the BH with passing through the outer CP (denoted by $\blacktriangle$) as also seen in Figure \ref{Fig3R}.  When the matter becomes super-sonic and if the preceding flow is compressed high enough then the shock is formed. Here the shock is formed with the combined effects of the centrifugal and pressure gradient force of the flow \citep{c89,ct95}. There are two main conditions for the occurrence of the shock in the flow: 1) The flow should be super-sonic at high enough distance from the BH, which can be allowed by the low AM distribution at the high $r$, and 2) The centrifugal force is also high enough in the inner region of the flow means $L_0$ should not be very low as seen in the PS Figure \ref{Fig4R}. Therefore, the shocked flow has larger value of the $\rout|_{min}\sim5\times10^3$ (possible minimum size of the shocked disk) as mentioned in Table \ref{tab:tab1}.
The other two solutions, $\be>1$ with the \hm gas at the AB have not sufficiently transported the AM in order the flow can become super-sonic far away from the BH, since the $\be$ (or $\Thob$) are less than the $\be$ (or $\Thob$) of the shock case for this particular $\rout$ (panels c and d). In case of $\be<1$, the AM transportation is comparatively inefficient around the AB, so that the AM deviates slowly from the the Keplerian AM distribution as seen in the panel a, and formed the ADAF solution with the \cm gas. 
Moreover, the variation of the $B$ is always greater than $1$ with the \hm gas for any kind of solution and unlike to the solutions with the \cm gas of the flow, the $B$ becomes greater than $1$ close to the BH as represented in the panel (d).  
We have zoomed the $\Theta$ profile of the solutions around the shock location ($\rs\sim7$) and represented in an inset of the panel c of the Figure. In the inset, the $\Theta$ is sharply raised at the shock location and the $\Theta$ is highest in the post-shock region. 
In this particular shock case, the $R\sim1.26$ is low and shock is weak, so the change in the $\Theta$ and $v$ is not much at the $\rs$ (Some more shock cases are represented in a Figure \ref{Fig6R}).
Thus the study of the nature of gas at the AB is very important, which can lead to change the inner disk structures so that may be the observables of the objects. As we have seen from this figure the differences of the $\lambda$ and $v$ is large in the flows of different kind of the solutions with only changing the $\Thob$ at the AB, for example, the difference in the $v$ of the shock and ADAF is more than two order of magnitude around the radial distance $r\sim10^2$ in the presented solutions. Thus these variations can alter a lot dynamical or viscous time scales therefore the effectiveness of the various radiative processes, which can change the non-thermal radiations of the BXBs and AGNs. Moreover, a bulk-motion Comptonization can also be more effective in the shocked solutions. 

\begin{figure}[ht]
\plotone{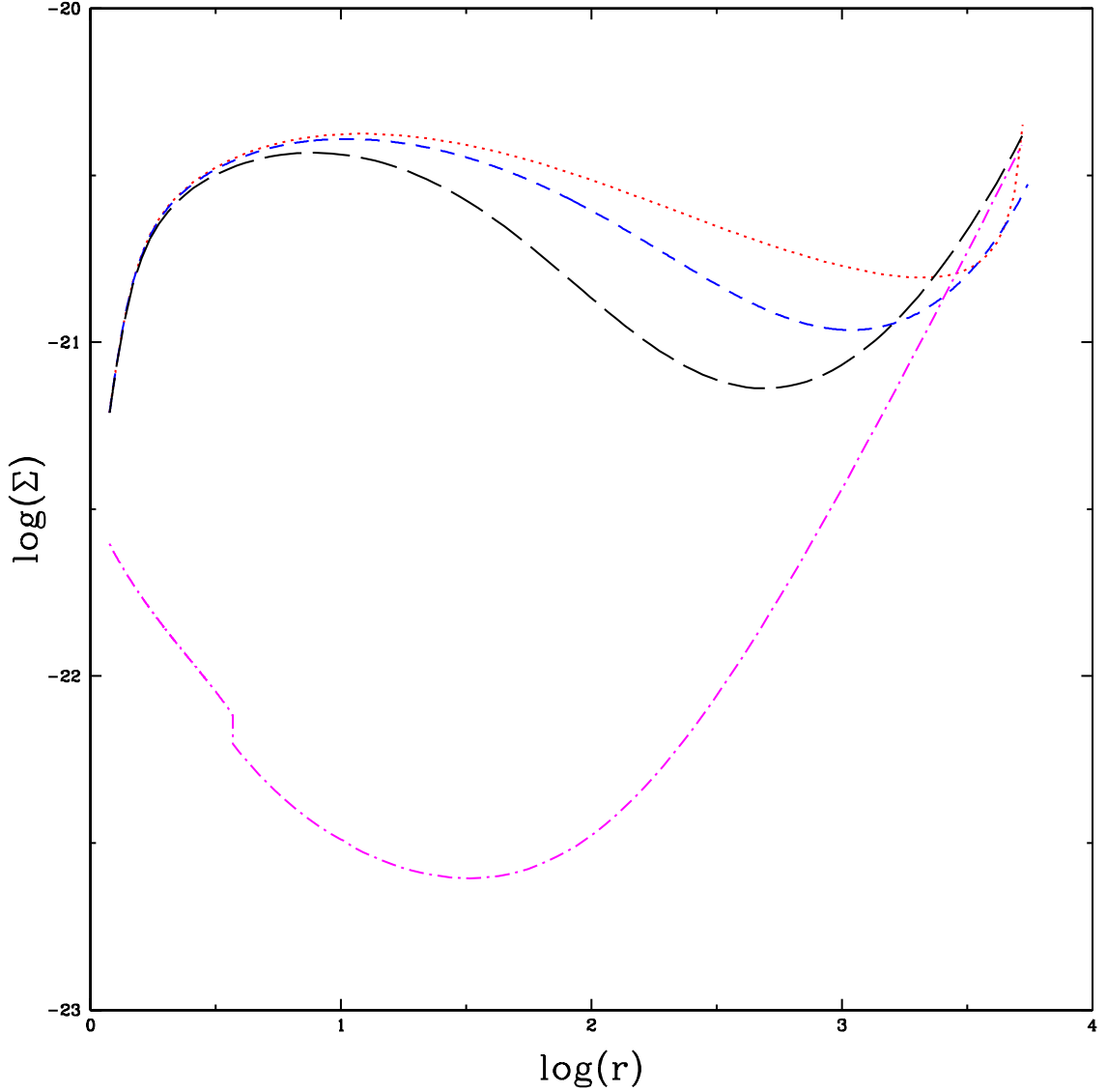}
\caption{Comparison of the profiles of the surface density $\Sigma$ are presented. The input parameters and line style of each solutions are same as in the previous Figure \ref{Fig5R}.
\label{Fig6Rs}}
\end{figure}
The expression of surface density of the flow is defined as $\Sigma=2\rho H=\dot{M}/(2\pi u^r r)$ from the equation (\ref{mdot.eq}). Here we choose $\dot{M}=0.1\dot{M}_{Edd}$, $\dot{M}_{Edd}=1.4\times 10^{17}(G/c^3)(\mbh/\sun)$ in the dimensionless form, and $\mbh=10\sun$. In the Figure \ref{Fig6Rs}, the $\Sigma$ is plotted corresponding to the solutions of Figure \ref{Fig5R}.
The $\Sigma$ profile is highest for the ADAF solution (dotted line) and lowest for the shocked solution (dashed-dotted line) with the same $\dot{M}$. Since the shock flow is fastest moving and ADAF flow is slowest moving as seen in the panel (b) of Figure \ref{Fig5R}. The $\Sigma$ is  decreasing in the outer disk in all the four cases. Due to the fast rise of the flow velocity is dominating over the rise of the temperature. 
In the middle region, the variation of the $\Sigma$ is mostly dominated by the rise of the $\Theta$ since $H\propto\Theta$.
In the inner region, the $\Sigma$ is decreasing due to combine effect of rising $v$ and decreasing $\Theta$ in the low $\be$ flow but in the shock flow, it is rising due to fast rise of the $\Theta$ (See panel c of the Figure \ref{Fig5R}). Thus the optical depth can also vary large with the variation of the $\be$. So the variations of the optical depth is not only dependent on the mass accretion rate and also dependent on the temperature of the gas at the AB, which can change the dynamics and the radiative processes of the flows.

\begin{figure}[ht]
\plotone{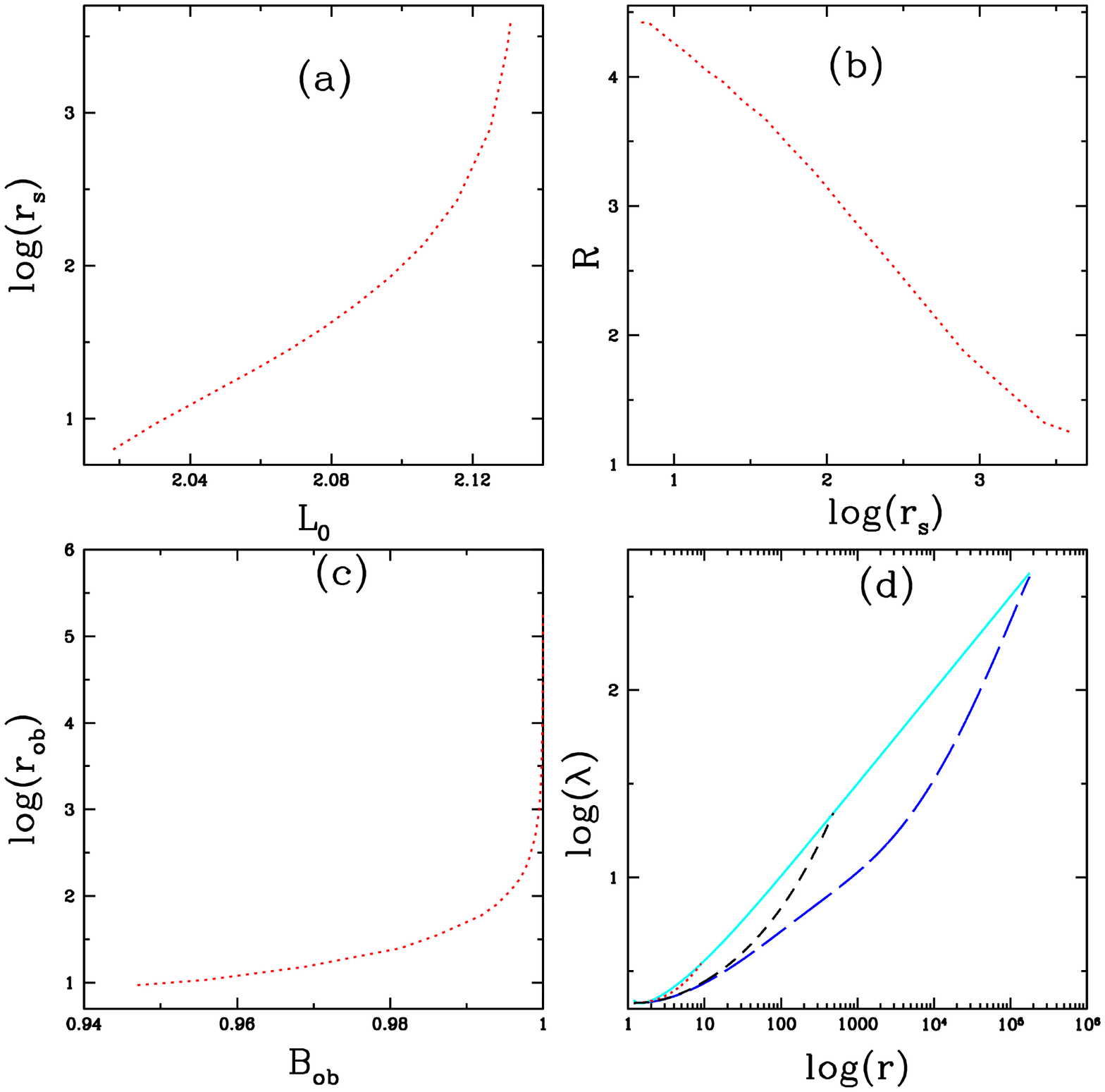}
\caption{Variations of the shock location $\rs$ with $L_0$ (panel a), compression ratio $R$ with $\rs$ (panel b), ADAF size $\rout$ with $\be$ (panel c) and $\lambda$ with $r$ (panel d) are presented. In the panel (a) and (b), the $\rs$ and corresponding $R$ are plotted for same $\be=1.00000055$ with different $L_0$. In the panel (d),  the dotted,  dashed and long-dashed curves are represented with three $\be=0.94630489,~ 0.99897261$, and $0.99999758$, respectively. The solid line (cyan color) is representing the Keplerian AM distribution. 
\label{Fig8R}}
\end{figure}
In Figure \ref{Fig8R}, we have presented some other properties of the ADAFs and the shock solutions, which can be helpful in the understanding of the various combinations of the inner structures/geometries of the disk with variations of the shock locations ($\rs$) and ADAFs size ($\rout$). 
So the combinations of the inner disk structures can be useful for the possible origin of the various observed variabilities, time scales of the states, and high energy radiations of the BXBs and the AGNs. In the panel (a) of the Figure \ref{Fig8R}, we have plotted the variations of the $\rs$ with $L_0$ and corresponding $R$ in the panel (b) for the fixed $\be\approx1.0000005$.  For the range of $L_0$, the shock location can move between $\sim6$ to $3800$ range, and compression ratio can vary between $\sim1.1$ to $4.5$. As we know that the higher $R$ means the matter is more compressed so the post shock flow can be more hotter and denser than the pre-shock flow. The higher $R$ in the post-shock flow can be more suitable for the inverse-Comptonization process \citep{ct95,mc10,kcm14} and this region can also produce the outflows \citep{gc13,kcm14,lckhr16,kc17,gbc20} due to the generated abrupt thermal gradient force in the post-shock region. 
In panels (c) and (d) of Figure \ref{Fig8R} are represented the variations of the $\rout$ with $\be$ for the ADAFs, and the $\lambda$ variations with $r$, for three values of the $\be$, respectively. The panel c shows that the outer AB of the ADAF is the higher for higher $\be$, and the ADAFs can not be found with $\be>1$.  The $\lambda$ distribution is higher for lower $\rout$ or $\be$ as seen in the panel (d). The dotted red curve becomes almost Keplerian AM distribution.
So the bulk velocity and temperature is also decreased with lower $\be$, but surface density is increased, which is shown in Figure \ref{Fig6R}. 

\begin{figure}[ht]
\plotone{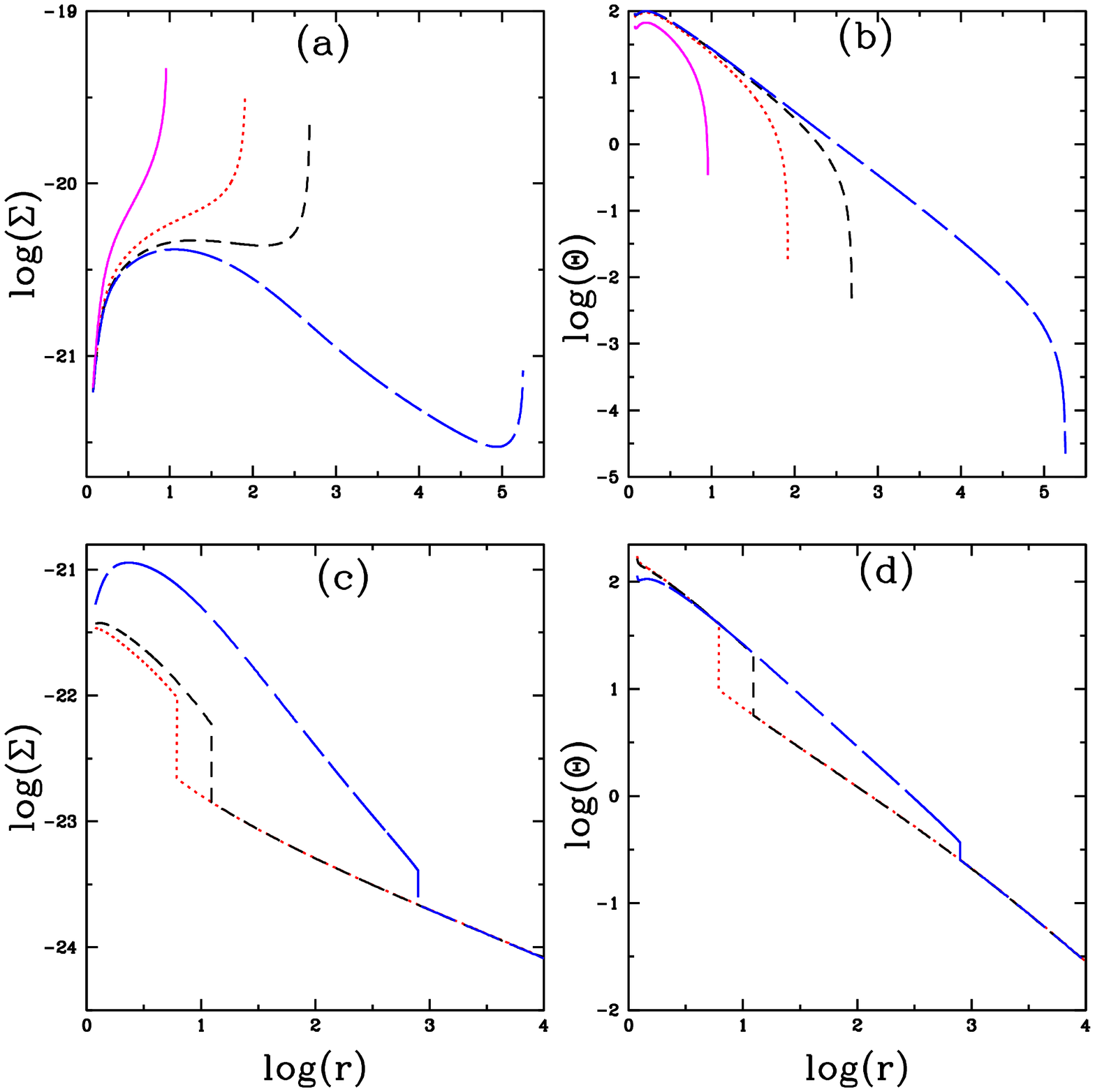}
\caption{The variations of the $\Sigma$ (Left panel) and $\Theta$ (Right panel) with radial distance $r$ for the ADAF and the shock solutions are presented. The ADAFs are plotted with different $\be$ or sizes. The shock solutions have different shock location with different $L_0$  and kept same $\be$ at the outer AB location. The surface density is calculated with $\dot{M}=0.1\dot{M}_{Edd}$.
\label{Fig6R}}
\end{figure}
\begin{figure*}
\centering
\gridline{\fig{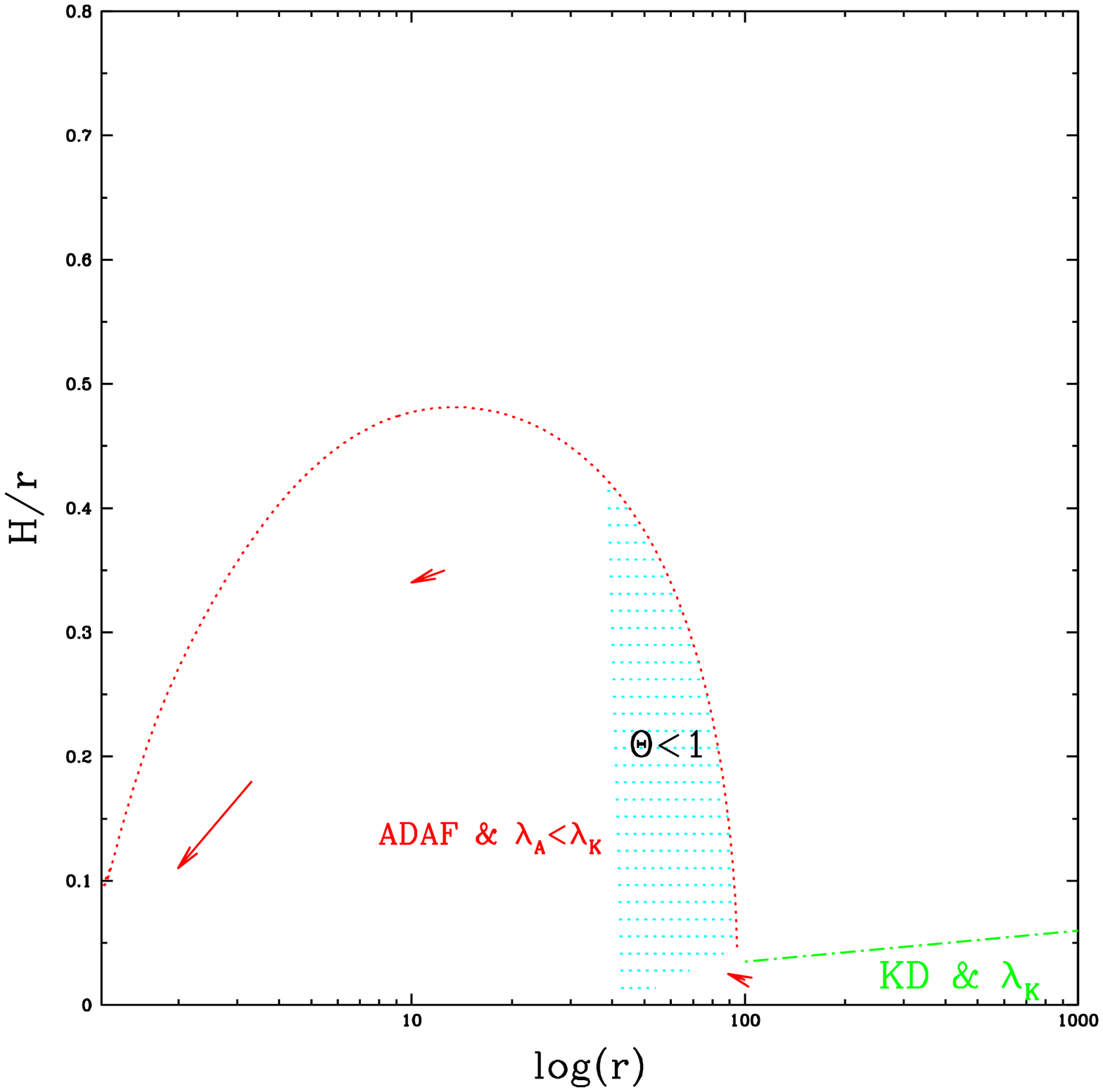}{0.5\textwidth}{(a)}
          \fig{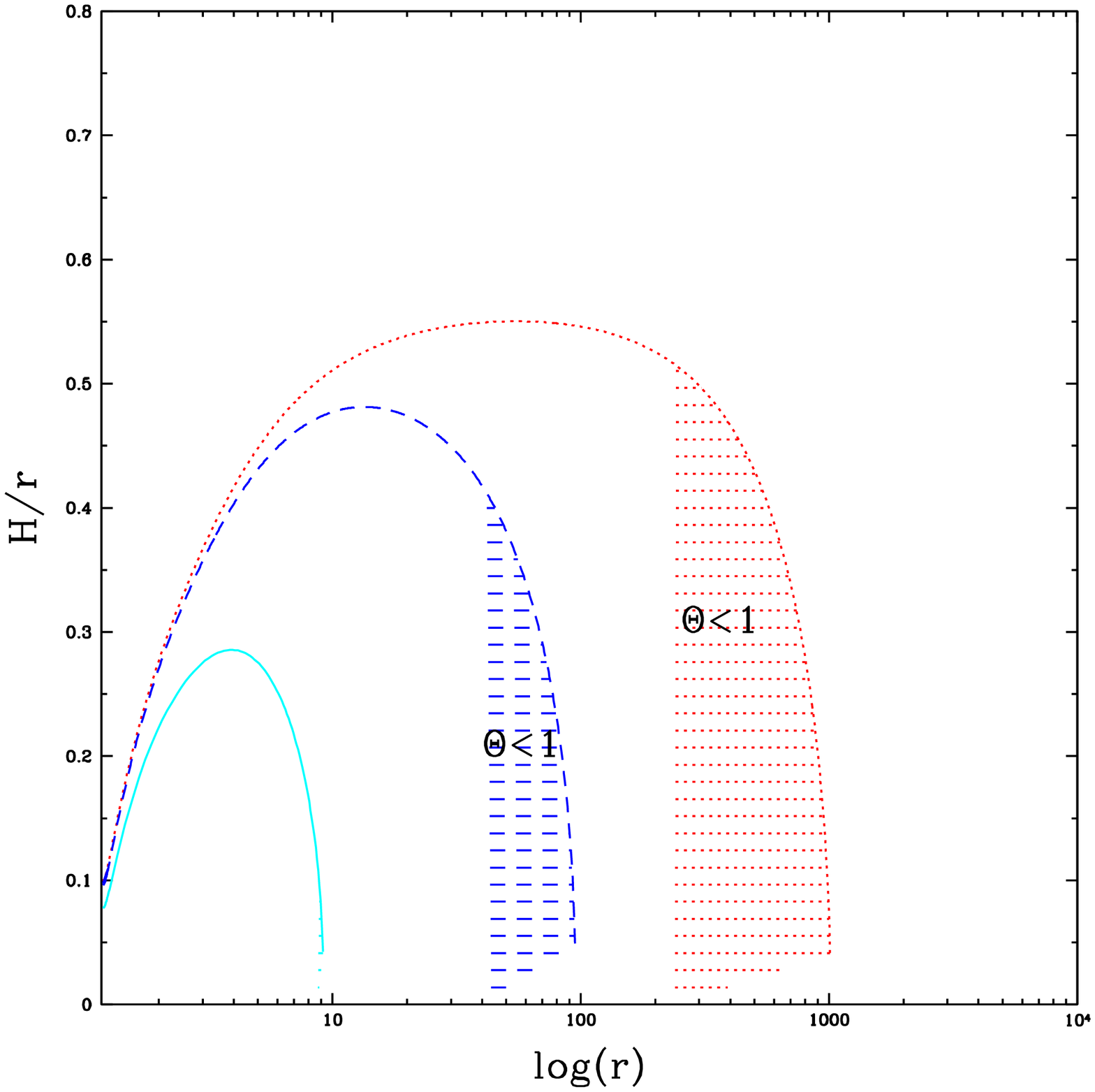}{0.5\textwidth}{(b)}
          }
\gridline{\fig{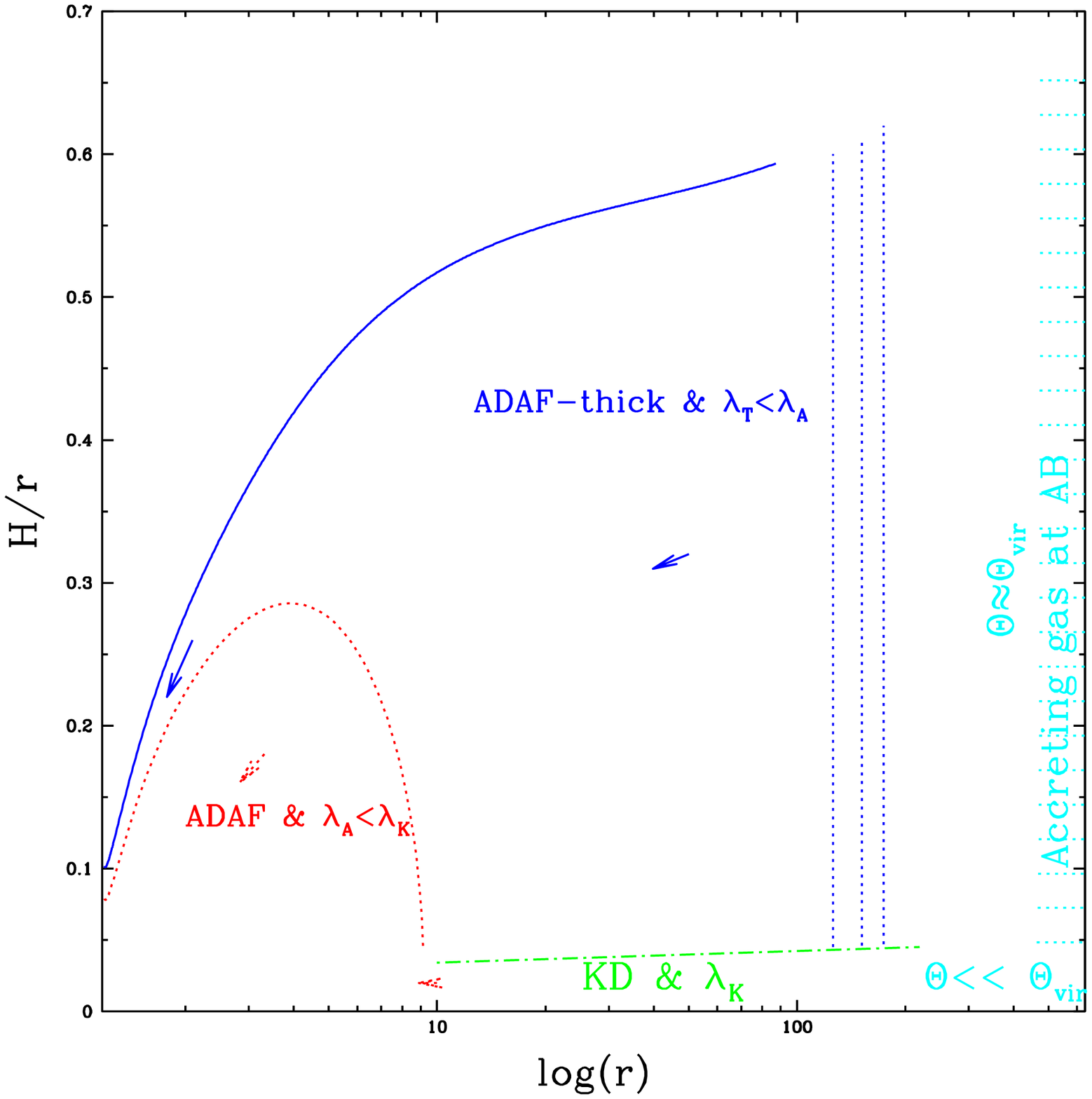}{0.5\textwidth}{(c)}
          \fig{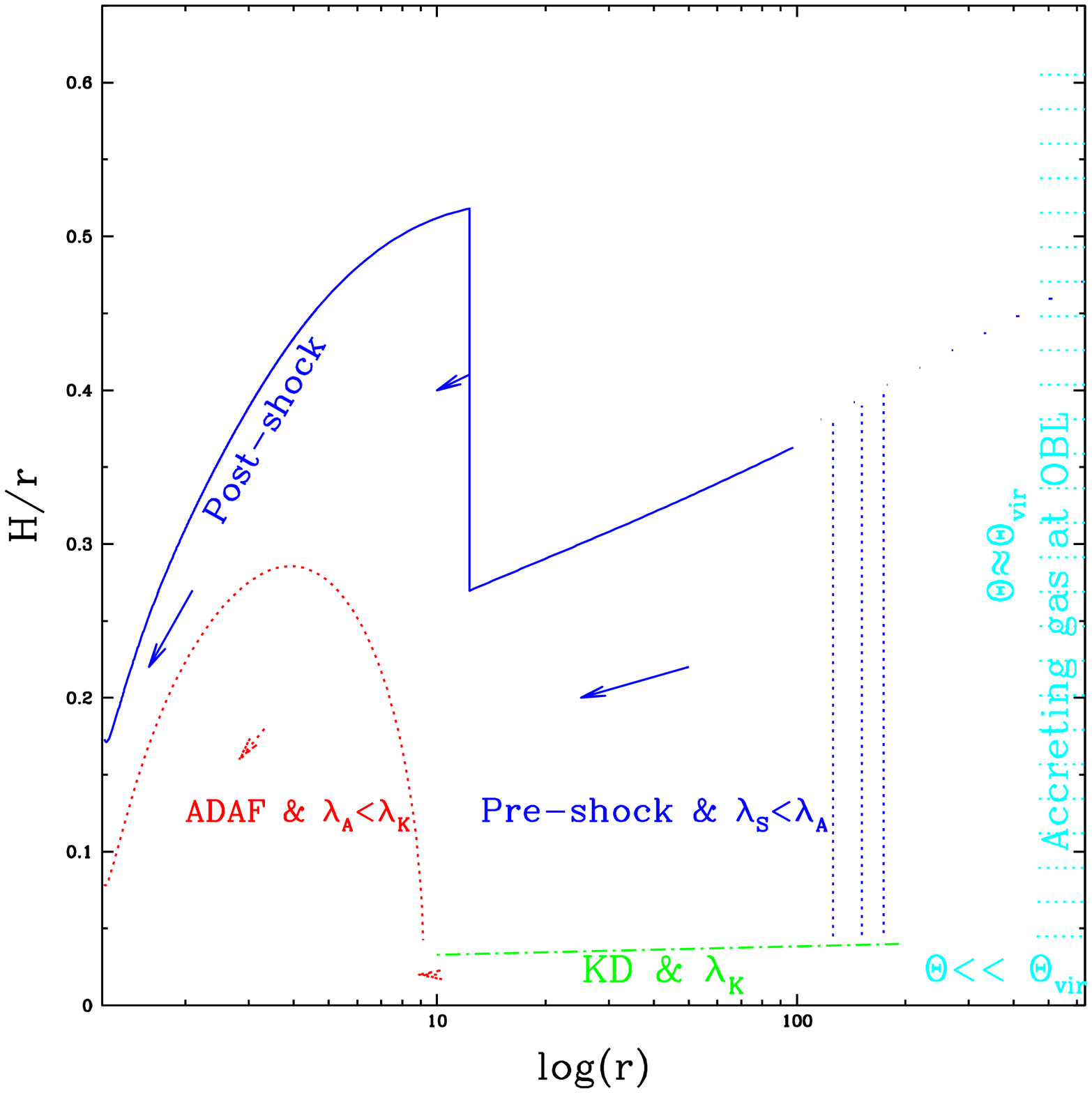}{0.5\textwidth}{(d)}
          }
\caption{The disk structures are presented with the estimated disk half-height ratio ($H/r$) versus r. In panel (a) represents height profile of the ADAF with assumed cool KD gas and in the panel (b) plotted height profile of the different ADAFs with different size or $B_{ob}$. In panel (c), the two height profile are plotted corresponding to the ADAF-thick/smooth flow (solid curve) and ADAF (dotted curve) in the inner region with assumed both mode (hot and cool) gas at the AB. In panel (d), the two height profiles are again plotted corresponding to the shock flow (solid curve) and ADAF  (dotted) with both mode of the accreting gas at the AB. Here, the cool KD is sketched with dash-dotted line (green color).
\label{Fig7R}}
\end{figure*}

For the detail understanding of the variations of surface density and temperature is represented in Figure \ref{Fig6R} for the different size of the ADAFs and shocked solutions with the different $\rs$. Here, we chose these two kind of solutions because the both have extreme nature in the density variation and their flow variables as seen in Figures (\ref{Fig5R} and \ref{Fig6Rs}). 
The $\Sigma$ profile is higher for the smaller size of the ADAFs in the panel (a) of Figure \ref{Fig6R}, since the bulk velocity is lower due to the higher AM distribution for the lower $\rout$ (panel d of Figure \ref{Fig8R}). Interestingly, the $\Sigma$ is higher in the outer disk (dotted and dashed curve solutions) and has temperature around $10^{10}$ to $10^8$K for the large range of disk region as seen in a panel (b) of Figure \ref{Fig6R}, and these behaviors can be found for the intermediate size of the ADAFs between $10<\rout\lsim10^3$. In this $\rout$ range, the $\Sigma$ profile is similar to the recent simulation results \citep{ikt20}.  The optical depth of this outer zone is roughly calculated around $\tau\sim10$ with Thompson scattering for $\dot{M}=0.01\dot{M}_{Edd}$. 
So we believed that this region/zone can be suitable for the bremsstrahlung emissivity and Compton scattering. Thus it can produce soft excess as observed in the AGNs and BXBs.  Thus the intermediate size ADAF disk can have two zones: one inner zone with high temperature $T=(\Theta m_ec^2/k)>10^{10}K$, which can give hard radiation and the other outer zone with the intermediate temperature range ($10^{10}K>T>10^8K$)  and high density can produce soft $X$-ray, so the related zone of the disk  is represented with the shaded region in Figure \ref{Fig7R}a and b. 
The solution with solid line curve (magenta color) is plotted for $\rout<10$, which has higher $\Sigma$ but the region of intermediate $\Theta$ is almost vanished and the other hand, the solution with long-dashed line curve (blue color) is plotted for the large ADAF disk ($>10^3$), which has lower $\Sigma$ in the intermediate range of $\Theta$ so this may not be suitable for soft $X-$ray emissivities.
The $\Sigma$ profile of the shock solutions also depend on the disk size $\rout$ but the inner region is largely affected by the $\rs$ as seen in a panel (c) of Figure \ref{Fig6R}. Interestingly, the $\Sigma$ and $\Theta$ (panel d) is sharply raised at the $\rs$, so the rise of the these quantities can make the post-shock flow region suitable for the generation of the very high energy non-thermal radiations. 
Generally, the $\Theta$ of the post-shock region is higher than the ADAFs, so it can produce more high energy photons.

With the studies of the above various kind of the advective hot accretion solutions, there are many kind of disk structures/geometries can be possible with changing the temperature of one mode or both mode of the accreting gases at the AB. 
We believed that the two-fluid motion theory can work in the accretion flow, so the two flows (for \hm and \cm gases) with the different densities and velocities can  coexist and the contact points of the both flows can be represented as an interface surface. The accretion flow with the \hm gas has low AM distribution than the accretion flow with the \cm gas at the AB (Figure \ref{Fig5R}).  As we understood that the high AM flow (low velocity) with high density is more stable and can settle around the equatorial plane and the low AM flow (high velocity) with low density can cover it as like the sand-switch geometry.
Thus, the radially and vertically stratified structures are presented in Figure \ref{Fig7R}, which has represented the interfaces between the outer cool KD and inner hot ADAF flow as the vertical interface at some radius and other the radial interface with variable angles between the inner hot ADAF and covered with even hotter shocked/ADAF-thick flow and the outer cool KD on the equator with covered by same hot flows. Moreover, the presence/absence/dominance of each accretion component depends on the accretion rates of the \hm and \cm gases.
In Figure (\ref{Fig7R}a), the vertical interface of the two flows has presented with the ADAF and the outer KD and we believed that this structure can be dominated with the \cm gas at the AB, which has been studied by many authors for the modeling of the non-thermal  and thermal spectrums of the BXBs and AGNs \citep[][and references therein]{nym95,emn97,gl00,mrl19,cbw20}.  
Here we have divided the intermediate size ADAFs into two-zones (the inner zone with $\Theta>1$ and the outer zone shaded region with $\Theta\lsim 1$) on the basis of $\Sigma$ variations and $\Theta$ of the flow as in the panel (a) and (b) of the Figure (\ref{Fig6R}). 
A variation of the shaded outer region of the ADAF is represented in the panel (b) of the figure \ref{Fig7R}, where shaded region is decreased with the decreasing of the ADAF size and it almost vanished for $\rout\sim10$. 
Previously, we have discussed that the inner zone can be much effective for the production of the non-thermal hard $X-$rays and the outer zone can be for soft $X-$ray/UV radiations in the intermediate size ADAFs. 
Since $\Sigma$ of ADAF is increasing very fast in outer zone with decreasing ADAF size (Figure \ref{Fig6R}a). 
So the outer shaded part of the intermediate size ADAFs can be suitable for the bremsstrahlung cooling or inverse-Comptonization process with optically slim medium, which can give soft excess as seen in the observations \citep{wxw20,swj17} and simulation \citep{ikt20}. 
Thus this theoretical study may be useful for the understanding of changing look AGNs with changing of the $\rout$ or $\be$ of the ADAFs. 
So, this possibility would be studied in detail with relevant radiative emissivities in the next future work. 

The radial interface situation between the non-advective KD and the shocked flow has already been studied for the modeling of the spectral states of the BHs by many authors \citep[][and references therein]{ct95,mc10,ndmc12,am20,bar20,pjc20}. As we have discussed that the accreted matter can have \hm and \cm gases on the AB, so on the basis of theoretical solutions, the accreting matter can form various kind of the flows. So we have proposed one more kind of the interfaces with the two advective flows in the inner part of the disk, which may exist between the ADAF and the shock or ADAF-thick hot flows as shown in the panels (c and d) of the Figure \ref{Fig7R}. So this could be the most general picture of the disk structure with both the radial and vertical interfaces with including the KD. 
This kind of the radial interface with the shocked flow and ADAF is also seen in the 2D accretion flow study by \cite{kg18}. 
The inner region of the shock flows have $\Thmx\sim180$ (not much depend on the $\be$) and the ADAFs have $\Thmx$ between $\sim 50$ to $100$, which depends on the $\rout$ or $\be$ as presented in Table \ref{tab:tab1}.  
At the shock transition, the gas temperature and density raised very sharply with a few times of the temperature and density of the pre-shock gas, respectively (Figure \ref{Fig6R}). So the post-shock region can be treated as the hot diluted corona and can suitable for the inverse-Comptonization of the soft $X$-ray photons, which can produce the hard $X$-ray and soft $\gamma$-ray. The density raised in the post-shock region can still keep the gas medium optically thin or slim, which is good for the Comptonization process.  
In panel c, the disk structure with the ADAF-thick flows can also have possibility to generate soft excess from their intermediate region of the flow.  Since these kind of the flows can be existed for the higher accretion rate than the other hot advective flows \citep{kc14} with the \hm accreting gas. So the dominant and weaker presence of this disk component may make the changing look AGNs.

We believed that these disk structures can give the possible explanations of the persistent/transient hard state behavior of the LMBXBs and HMBXBs. Since the BH of HMBXB is mostly fed by the stellar winds. The winds are hotter ($\Theta\sim\thvir$), 
so the BH can mostly be fed by the \hm gas, which can produce the shocks in the inner part of the disk or ADAF-thick flow and the disk structure can look like Figure \ref{Fig7R} (panels c and d). The flow with the \hm gas is dynamically faster than the \cm flow (Figure \ref{Fig5R}b). So there is more possible to keep object in the hard state or bring in the hard state in the shorter time scales with supplying the gas in the inner part of the disk. On the other hand, the LMBXBs can mostly be fed by the \cm gas (Roche-lobe flow), which can form the KD or ADAFs or both, like structure of Figure {\ref{Fig7R}a}. The flow with the \cm gas has the larger dynamical or viscous time scale (may depends on the KD flow) than the \hm gas, so the inner part of the disk can take much time to fill with the gas in the LMBXBs, so it spend most of the time in the quiescence state. Thus the hard state of the HMBXBs are more persistent than the LMBXBs.     
Contrary to the conditions of the HMBXB, if the disk size is very large around the BH in the LMBXB then it can also possible that the source can spend more time in the hard state, for example, GRS1915+105 is the LMBXB and frequently found in the hard-state \citep{nb18}. 
Since the large disk has the large amount of the accreting gas, and the outer part of the disk can work as a gas reservoir, so the BH can feed with more continuity.    
Now we can believe that the similar analogy can also be found in the persistent bright AGNs with the large disk, and the AGNs also have many kind of the gas sources. So if we can use temperature ($\Thob$) as a one more parameter with the accretion rates then the understanding of short/long time variabilities, time scales of the various typical spectral states can be understood with the tuning of the both mode gases.  
Although, we are expecting some more studies with the theoretical calculations regarding these kind of objects in the future with incursion of relevant radiative emissivities.  Moreover, for the detail understanding of these scenarios, we will also need to do simulation studies with large length scale and time dependent conditions with two-mode gas inflow at the AB.

Moreover, we can understand that the high AM flow can more stable than the low AM flow with dynamics of the flow, like, the KD is more stable than the ADAF and the ADAF is more stable than the shock solution (or the flow with the \hm gas), since the AM relation is like $\lambda_K>\lambda_A>\lambda_S$, respectively (Figures \ref{Fig7R} and \ref{Fig5R}a). 
Since high AM flows reside close to the equatorial plane, which can make more stable orbits. 
Thus the shock solution can have more chance to become unstable with small amount of the perturbations, which can break the vertical equilibrium, so they can generate the outflows or break the local radial equilibrium (due to change in the dynamical, viscous or cooling time scales, or intermittent accretion rate) can make unstable post-shock or inner part of the disk. Which can introduce variabilities in the non-thermal radiations \citep{ct95,dcnm14,lckhr16}. Thus the shocked flows can have more exotic features than the ADAFs and the KD flows.

\subsection{Physical or Unphysical solution?}\label{subsec:ups}
\begin{figure}[ht]
\plotone{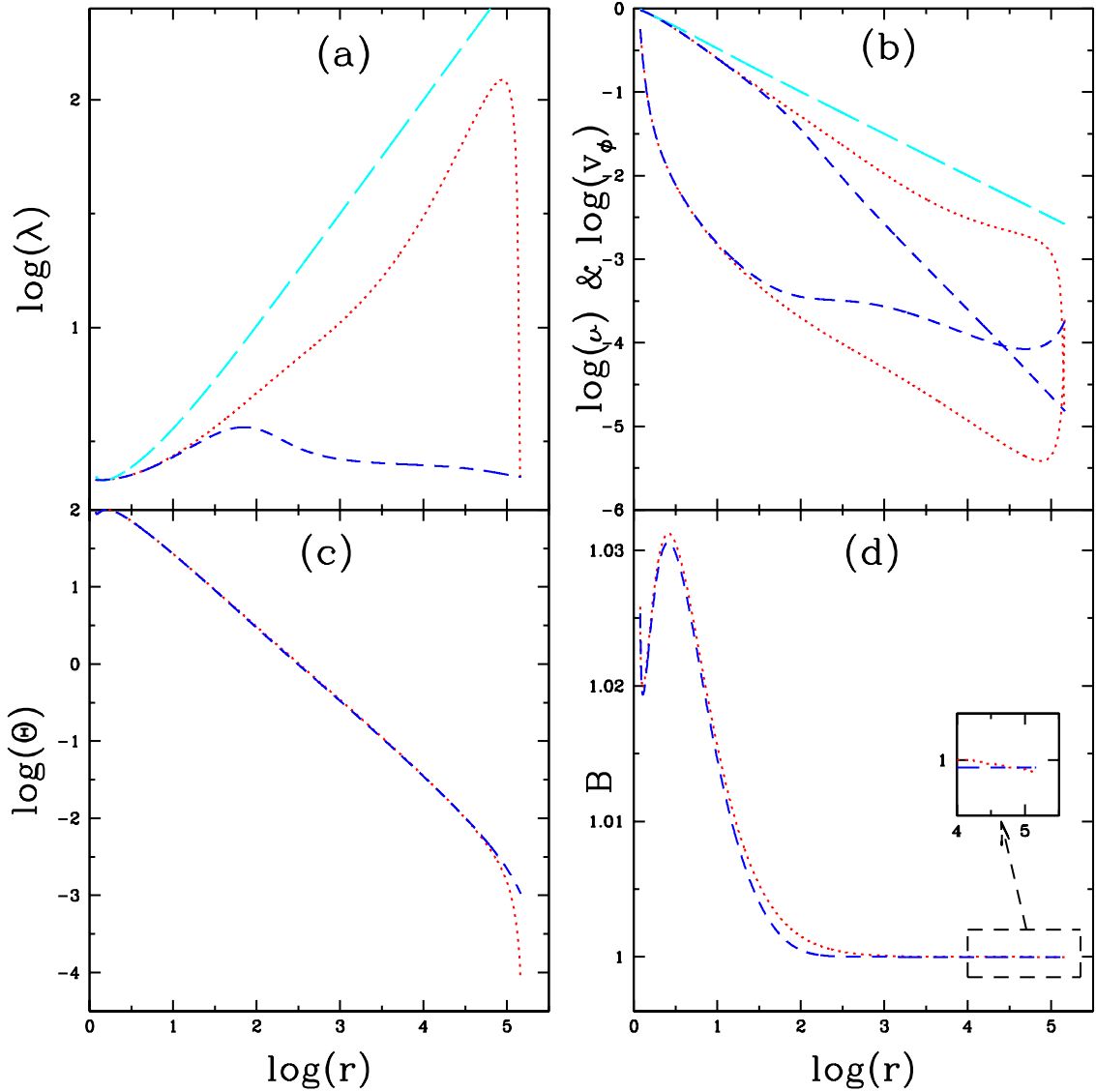}
\caption{Variations of flow variable $\lambda$ (panel a), $\vr$ and $\vp$ (panel b), $\Theta$ (panel c) and $B$ (panel d) with $log(r)$ are presented  with different $\be=0.999993$ (dotted curve) and $0.999996$ (dashed curve). The long-dashed curves in the panel a and b are representing the Keplerian AM and Keplerian velocity ($\vk$) distributions, respectively. In the panel b, the upper curves with close to $\vk$ are representing $\vp$ variations and the lower curves are representing $\vr$ of the flows.
\label{Fig9R}}
\end{figure}
The outer ABCs can be $\be<1$ ($\Thob<<\thvir$) with $\lmdob\sim0(<<\lmk)$ and $\vr\gsim v_\phi(<<\vk)$ at the AB location\replaced{of}{, and the corresponding ABCs, the} two solutions with two different $\be$ are presented in Figure \ref{Fig9R}. The parameter region of these kind of solutions is located below the long-dashed line ($\be=1$) with denoted as $\alpha-$type in the PS Figure \ref{Fig4R}.
Here $\vk$ is the Keplerian velocity. These kind of solutions are treated as the unphysical solutions in many studies \citep{c89,lgy99,ck16,kc15,kc14,kc17}. Since they are making close topology multivalued solutions (inverted $\alpha$- type) with the spiral path due to the second/middle $O-$type CP, which makes unusual behavior of the flow variables from the understanding of the typical accretion flows and most of the solutions are the short range size, especially with lower $L_0$ and $\be$ but they can become the physical when connected with some kind of the dissipative shocks via the outer CP flow \citep[][and references therein]{kc13}. 
Since nothing is static in the Universe so there is a possibility that the gas cloud can have some initial local velocity, which may higher than the respective local $\vp(<<\vk)$ at the AB when the gas is captured by the BH. After capturing of the gas, the bulk radial velocity of the gas cloud can be decreased or increased may depend on the re-orientation of local thermal motion of the gas and motion of the gas cloud, and initial AM distribution with respect to the accretion center. Although the decreasing bulk velocity seems unphysical in accretion around the AB as in panel b (Figure \ref{Fig9R}) but the above stated ABCs can be possible.  So the possible reason of this may be that the AM transportation is highly inefficient in these cases. 
The highly inefficient AM transportation means the outward viscous transportation process is too weak (maybe due to very low $\Thob$ and $\lmdob$ at the AB) than the other local dynamics (maybe some unknown process!, which becomes efficient to increase the $\lambda$) or no continuous gas medium to transfer the AM outward then the AM can be gained in the outer part of the gas cloud as moved in (as see variation of $\lambda$ in the panel a) due to the central gravity.
Thus the $v_\phi$ is raised very fast than the $\vk$ but still less than the $\vk$ values (panel b), which continue the inward motion with the slowing down of the radial velocity ($v$) around the AB. 
When the disk is formed around the object with sufficient combine effect of $\rho$ and $\Theta$ then the gas flow variables behaved well as like usual accretion flow, which is also seen in the inner part of the flow.
In the Figure \ref{Fig9R}, we have plotted two solutions with different $\be$ at the AB as the zoomed outer part is represented in the inset of the panel d. Although the difference of the $\be$ is not much but the differences in the variations of the flow variables is a lot, especially the $v$ and $\vp$ or $\lambda$ of both the solutions in the panels a and b.  In the panel b, the values of the $v$ and $\vp$ are same for the both cases at the outer $\rout$ but the $\Thob$ is different around one order of magnitude in the panel c. It indicates that the initial local temperature of the accreting gas is very important component for the accretion flows. 
Although, these situations are very difficult to understand with the present theoretical model, but it can be understood by way of art of simulation with suitable mathematics and physics.
These solutions can be helpful to understand the disk formation in the young stellar objects (YSOs) and transient disk formation in the tidal disruptive events (TDEs) with the following theoretical illustrations/demonstrations.  

{\it Illustration-1:} A gas cloud of a star forming region has mostly the local motion and no bulk motion of the gas before satisfying the Jean's criteria. When the cloud collapsed and start accreting surrounding gas then bulk motion is developed. The direction of this bulk motion can be radial and azimuthal (probably due to re-orientation of the local velocity distributions of the gas in presence of the gravity of the central object). Thus the disk is formed due to azimuthal velocity component ($v_\phi$). Here the azimuthal velocity is generated with respect to the accreting object (maybe similar to the AM in the panel a) and therefore the AM. So these solutions (Figure \ref{Fig9R}) can represent the YSOs disk and transient disk of the TDEs.  
Thus, we believed that this kind of solutions are also physical and can represent the accretion flow with the large outer AB. 

{\it Illustration-2:} Suppose, an solid object/gaseous cloud is come into the influence of gravitational sphere of the central object with the sub-Keplerian AM at the outer AB. Here, the accreting object does not has continuous medium to transport the AM outward, but  the object has the sub-Keplerian AM, so it moves inward until to obtain the Keplerian AM or orbit ($\vp\rightarrow\vk$ with $\vr\rightarrow 0$), so here $v$ is decreased (maybe similar to the $v$ in the panel b). However, the AM will be constant in case of the solid object and the AM can be redistributed in the case of the large gaseous cloud. So there is a possibility that some part (mostly outer part due to the AM transportation from the inner to the outer part) of the large cloud can gain the AM as move in, thus it can make the transient or persistent disk.
This situation may happen with the tidally disruptive gas clouds and may form the transient disk close to the central object.    

{We believed that the gravitational and centrifugal forces are the omnipotent forces in our Universe, and the centrifugal force (pseudo-force) is a like mirror image force and generated due to the gravity pull. So, if a sufficiently large object come into the gravitational influence of a heavy object then some amount of the centrifugal force must be originated with respect to the accretion center due to the 
 movement inertia of the accreting object or presence of the space-time curvatures around the outer boundary of the other astrophysical objects. 
Thus the solid accreting object can make the elliptical or spiral trajectories and the gaseous accreting object can make disk or disk-like structure.}

\section{SUMMARY AND DISCUSSION}
In this study we have investigated all the possible advective transonic solutions with the detailed computational analysis (presented in the section \ref{sec:nmethod}) of the GRHD equations around the Kerr BH and represented them in a single plane of the PS in Figure \ref{Fig4R}. We have also discussed the possible physical perspectives of them.  
We have found that the AM transportation is more effective in the outer part of the disk with increasing the temperature of the accreting gas at the outer AB. Which changed the viscous and dynamical time scale of the flow so it can change the radiative emissivities or excite the variabilities in the flow. We also identified the two-type of the accreting gas at the AB with their local energies of the gases, and the variations of the AB locations ($\rout$) as seen in the Figure \ref{Fig2R}a and named the gas with $\be<1$ is as the \cm gas ($\Thob\ll\thvir$) and the other $\be>1$ as the \hm gas ($\Thob\lsim\thvir$).  
On the basis of the obtained solutions with two-type of the gases and their nature, we have made some theoretical points as well as the possible explanations of the observed properties of the accreting astrophysical objects. Which are the following: 

(i) Here, we have explored in very detail the \hm and \cm accreting gases at the AB on the basis of nature of the advective disk solutions \footnote{The similar kind of study has been done with pseudo-Newtonian geometry around a non-rotating BH by \cite{y99}. His computational and physical analysis was limited with three kinds of the advective solutions}. 
The \hm gas is more effectively transported the AM outward around the AB than the \cm gas, which can lead to the origin of the different kind of the accretion solutions. 
The \cm accreting gas can generate ADAFs and two CP multivalued solutions. The possible dominating sources of this kind of accreting gas may be the gas of the KD, Roche-lobe over flow, ISM, torus gas, cold clumps in the broad region of the galaxy and so on.  
The \hm accreting gas can generate MCP or single CP solutions, like, the shock solutions, smooth solutions (or ADAF-thick) and Bondi type solutions. The possible dominating sources of this kind of the accreting gas may be failed winds from the ADAF disk or the KD, stellar winds, hot gas clouds in the broad region of the AGNs and so on.

(ii) We found that the gas density of the flow is changed a lot with changing the temperature of the accreting gas at the AB (Figures \ref{Fig6Rs} and \ref{Fig6R}), so the optical depth of the flow can also change. Which can alter the non-thermal radiative emissivities of the flow. Thus the radiative emissivities are not only dependent on the accretion rate but can also dependent on the initial temperature of the accreting gas at the AB. So It can be useful to study the temperature of the accreting gases at the outer AB with the observational studies, which may help in the understanding of the behaviors of the accreting BHs.

(iii) When we compared the most popular solutions, the shocked flow and the ADAFs then found some interesting results as follows: 
{\it I}) The ADAFs has the higher AM distribution, so they could be more stable with the external perturbations. So any change in this kind of flow could be more smoother and slower. 
{\it II}) The shock solutions are dynamically faster, so it can faster change the spectral-states of the BXBs and AGNs and,  
{\it III}) The post-shock flow is always found with the higher temperature than the ADAF (Figures \ref{Fig5R} and \ref{Fig6R}). 
So the puffed-up post-shock region can act as the hot corona, which can give the more higher energy photons.
Moreover, the shock flow can easily become unstable or oscillatory in the post-shock region with the small amount of the mis-match (or change) between the local dynamics (time scales) or some kind of perturbations, like, the failed winds or intermittent inflow rates, which can be suitable for the observed variabilities in the spectrum.

(iv) We proposed the maximum three-component accretion flows in the disk geometry with including the KD flow (Figure \ref{Fig7R}), which can be the helpful in explanation of all the states of the BXBs and the AGNs with the tuning of the accretion rates of both mode gases. Out of the three, the two components are the advective in nature and are investigated here. The advective components can be divided into many zones on the basis of their temperatures and variations of the density of the flow, like, the intermediate size ADAFs can be divided into the inner part as the hot flow ($\Theta>1$)  and the outer part as the intermediate temperature ($\Theta\lsim1$) with the higher density (Figure \ref{Fig6R}), which can be the optically slim and the cooling effective. 
Thus we are expecting that the outer zone of the ADAF can help to understand the origin of the soft excess with changing of the ADAF size in the AGNs and BXBs. Similarly, we believed that the ADAF-thick solutions can also generate soft excess component with changing the accretion rate of the \hm gas, which can help in the understanding of the changing look AGNs.
Since the accretion flow with the \hm gas is much faster, so it can help in the study of some appeared and disappeared properties of the AGNs. 

(v) In the present accretion study, we found that the temperature at the AB can be one more parameter as the mass accretion rate. The temperature parameter can help to make more suitable for the understanding of the origin and modeling of the various properties of the accreting objects, especially, the time scales of the variabilities, and the time scales of the spectral states in the BXBs and AGNs. Since we believed that the timings or time scales of the states around the BHs depend on the size of the systems as well as the accretion rates of the both mode gases. 
For instance, the LMBXBs mostly exhibit as the transient hard $X-$ ray source than the HMBXBs. So the possible explanation of this behavior could be  that the BHs of the LMBXBs can be mostly fed by the \cm accreting gas, so the disk structure can mostly similar to those in Figure (\ref{Fig7R}a). Therefore, the inner disk of the LMBXB takes large time to fill with gas again after the out-burst since the flow with the \cm gas is dynamically slow and mostly depend on the time scale of the KD flow. 
On the other hand, the BHs of the HMBXBs can be mostly fed by the \hm gas, so the disk structure can be mostly similar to those in Figure (\ref{Fig7R}c and d). The flow with the \hm gas is faster and hotter, which can fill the inner disk faster after the out-burst when the gas is available. 
The similar analogy can be extended for the case of active/inactive AGNs and changing look AGNs. 
So it will be very interesting to do the detail spectral modeling of the objects with the theoretical and simulation studies by changing the temperature of the both kinds of the accreting gases as well as their accretion rates at the AB in the future studies.

(vi) These kinds of the theoretical studies can be very helpful to encourage the numerical simulation studies with using various ABCs for the large length scales, in order to generate the various kinds of the solutions with the two-mode accreting gases at the AB.

(vii) The full $r-\phi$ component of the viscous shear tensor is the more effective close to the BH, unlike the other simplified forms of it as used in the previous studies. Since the local energy ($B$) of the flow increases very fast close to the BH as seen in Figures \ref{Fig5R}d and \ref{Fig9R}d. it can be due to the radial velocity derivative in the viscous shear tensor, which is more effective close to the BHs.

(viii) We found that some two CPs solutions with the outer ABCs, $\be<1 (\Thob\ll\thvir)$, $\lmdob\ll\lmk$ and $\vr\gsim\vk$ for the large $\rout$ as presented in Figure \ref{Fig9R}.  The variations of the flow variables are unusual as we understood for the typical accretion flows, but these kinds of solutions can represent the transient disk of the TDEs and the disk formation in the YSOs.

(ix) Recently, the production of the neutrino flux has been investigated with the inner-outer blob model by \cite{xlw20}. Here they assumed that the inner blob is surrounded by a hypothetical $X-$ray producing hot corona of the accretion disk, and also vindicated that the inner blob can be formed with standing shocks in the jet base under the certain dissipative process. Now we believed that the hypothetical hot corona can also be formed with the standing shocks in the accretion flow as seen in our study. If so, the theoretical picture of this neutrino model with the standing shock in the jet base as the inner blob and the puffed-up post-shock region in the inner disk as the surrounding hot corona has been already investigated in the study of the accretion-ejection flows by \cite{kc17}.

\added{(x) Interestingly, \cite{drd20} have investigated successful, restricted-successful and hard-only outburst cases of a source GX 339-4 (see Figure 9 of the cited paper). The hard-only outbursts are not succeeded by the high soft-state and having low luminosity states only. The successful outbursts are following typical q-diagram on the hardness-intensity plot. The restricted-successful outbursts are also like the hard-only outburst but their radio and $X-$ray luminosities correlation same as the successful outbursts. Now these situations of a source may not be completely understood by the ADAF and KD (produces soft $X-$rays) coupling model only (Figure \ref{Fig7R}a), since the size and accretion rate of the ADAF is dependent on the KD flow \citep{emn97,kg19}. So these can be understood by the accretion flow which is independent of the KD flow, 
like, the \hm gas accretion flow. 
Since we believed that the \cm accretion rate can be low or not sufficient so the source did not enter into the high soft-state, and the inner region of the disk can be mostly filled by the \hm gas, which makes the sufficient material build up in the disk for having the outburst in the restricted-successful or  hard-only outburst cases. 
So this could be a way to understand these kinds of the behaviors of the BXBs. 
Thus here seems to favor the two-mode gas accretion flows at the AB. 
}

In the present study, we have investigated the accretion flows with a single values of the spin parameter, composition parameter and viscosity parameter but there are many things need to be explored in this model equations, like, the effects of the variations of the composition parameter, the spin parameter of the BH, and the viscosity parameter with/without including the radiative emissivities on the accretion flows, and the self-consistent study of stationary inflow-outflow solutions. So we have planned these studies to communicate in the future with next few papers.

{\it Improvements required:}
1. We have calculated the flow variables on the equatorial plane, so we need to improve it for the off-equatorial plane as well for the better understanding of the model solutions. Since the shocks can also formed above the equatorial plane \replaced{\citep{kg18,ftt07}}{as seen in the viscous 2D HD flow \citep{kg18}, and MHD flow \citep{ftt07}}. \added{So, we will be expecting that the study of the 2D flow with relevant radiative emissivities can give the detailed vertical structure of the accretion disk.}

2. In the present study, we changed only the temperature of the inflow gas at the outer AB. As we knew that 
the advective solutions depend on the AM distribution, which depends on the viscosity of the flow and the outer ABCs. Since, the viscous stress tensor is basically a pressure tensor 
and mostly depends on the $\rho$ and $\Theta$.  Therefore, the nature of advective solutions ultimately depend on the initial temperature and initial density of the gas (or inflow accretion rates).   
So hopefully, in the future studies with considering relevant radiative emissivities, we can present both the radiative efficient and inefficient solutions of the disk with tuning of the initial temperature and the density. Moreover, the AM distribution is also affected by the mass loss from the disk ({Kumar, Gu, \& Yuan under preparation)}

3. We also need to repeat same kind of the study with the two-temperature flow dynamics. Since the proton and electron temperatures can be different in the low accretion rate flows.

4. In the present study, we assumed that the KD is already present as a cool gas source for the ADAF solutions. So we need to understand the physics and the outer ABCs for the generation of the KD or KD-like flow. Moreover, we believed that since the time of evolution of this non-advective flow is very large, so it can be formed gradually from the cold gas at the AB. We also believed that the KD is most persistent disk flow and the inner part can change into the ADAF solutions due to some physical processes, like, the internal instabilities or external perturbations \cite[see][and references therein]{kg19}. Thus this important part also needs to explore in the future studies. \added{However, the transition of the optically thick to thin disk (or cool KD to hot sub-Keplerian flow) along the radial direction with the vertically averaged disk has been shown for the continuous  transonic  
 accretion flows with the suitable cooling/heating rates formulae in the pseudo-Newtonian regime by the some authors \citep{wl91,h96,abi06}. So it would be good to do detail study of the accretion flow with the continuous thin/thick cooling rate formulae.}

\added{Here we have rigorously investigated the transonic steady-state solutions in the HD regime with the help of computational and mathematical details, so it would be good to test those solutions in the MHD regime with considering some physical processes, like, the ohmic heating, Hall effect, Coulomb coupling for the exchange of the heat between ions and electrons, coolings  and so on. Since there is a possibilities that all the advective solutions (specially the ADAFs, and the shocks as discussed in the subsection \ref{subsec:ps}) may not be generated with dominance or incursion of the some physical processes, like, the ADAFs can not appear because of the electron heating by the magnetic reconnection \citep{bl97,bl01}, and the standing shocks can not form due to the dominance of the heatings or coolings \citep[high viscosity or high accretion rate or both][]{kc14,abi06}  or mass-loss or magnetic field in the flow. Recently, \cite{gbc20} have shown the existence of the quasi-stationary shocks in the MHD simulation with the outflowing disk, unlike to the steady shocks have found in the HD simulation with low viscosity \citep{lckhr16}, and in the steady-state MHD flow \citep[][references therein]{ftt07}.
}

The BH feeding mechanisms with the nature of the accreting gases, and the cause of viscosity with the low and high temperatures (or densities) flow are need to be explored more carefully by the observational and the theoretical studies, respectively. 
Final conclusion of our present study is that
all the accretion solutions/models, like, the shock solution, the smooth solutions (ADAF and ADAF-thick), the Bondi like flow and the KD can be a part of any kind of the accreting objects, and these can depend on the abundance, and nature (temperature and maybe composition) of the accreting gas at the AB. 
We strongly believed that the two mode accreting gas can well represent the observed phenomena of the accreting BHs. In order to understand the detailed effects of these gases together, so we need to do simulations with two-type of gas inputs on/off the equatorial plane. Hopefully, we can do so in the future.
\acknowledgments
We thank the referee for his useful suggestions, which improved the contains of our manuscript.
This work is supported by National Natural Science Foundation of China (Grant No. 11725312, 11421303). RK is supported by CAS President’s International Fellowship Initiative (PIFI). 
RK thanks Dr. Sheng Z.-F. for helpful discussions.

\end{document}